\documentclass[aps,prc,twocolumn,groupedaddress,showpacs]{revtex4-1}

\usepackage{graphics}
\usepackage{amsmath}
\usepackage[normalem]{ulem}
\usepackage{comment}
\usepackage{epsfig}
\usepackage{color}  
\usepackage{hyperref}

\def\nuc#1#2{\relax\ifmmode{}^{#1}{\protect\text{#2}}\else${}^{#1}$#2\fi}
\newcommand{\etal}{\textit{et al.~}}

\newcommand{\be}{\begin{eqnarray}}
\newcommand{\ee}{\end{eqnarray}}
\newcommand{\vecr}{\vec{r}}

\newcommand{\dent}[3]{\rho_{#1,#2 \rightarrow #3 }}

\begin{document}


\title{Semi-microscopic folding model for the description of two-body halo nuclei} 



\author{J.~A. Lay}
\email{lay@us.es}
\altaffiliation{Present address: Dipartimento di Fisica e Astronomia, Universit\`{a} di Padova, I-35131 Padova, Italy \\ 
 Istituto Nazionale di Fisica Nucleare, Sezione di Padova, I-35131 Padova, Italy}
\affiliation{Departamento de FAMN, Facultad de F\'{\i}sica, Universidad de Sevilla, Apdo.~1065, E-41080 Sevilla, Spain}

\author{A.~M.\ Moro}
\email{moro@us.es}
\affiliation{Departamento de FAMN, Facultad de F\'{\i}sica, Universidad de Sevilla, Apdo.~1065, E-41080 Sevilla, Spain}

\author{J.~M.\ Arias}
\email{ariasc@us.es}
\affiliation{Departamento de FAMN, Facultad de F\'{\i}sica, Universidad de Sevilla, Apdo.~1065, E-41080 Sevilla, Spain}

\author{Y.\ Kanada-En'yo}
\email{yenyo@ruby.scphys.kyoto-u.ac.jp}
\affiliation{Yukawa Institute for Theoretical Physics, Kyoto University, Kyoto, Kyoto 606-8502, Japan}


\date{\today}


\begin{abstract} 
One-neutron halo nuclei, composed by a weakly-bound particle coupled to a
\textit{core} nucleus, are studied within a particle-plus-\textit{core} model. A
 semi-microscopic method to generate the  two-body Hamiltonian of such
a system, including \textit{core}
excitation, is proposed. The method consists in generating the spin-independent
part of the valence-\textit{core} interaction using a single-folding
procedure, convoluting a realistic nucleon-nucleon (NN) interaction
with the \textit{core} transition densities. The latter are
calculated with the Antisymetrized Molecular Dynamics (AMD)
method. The prescription is applied to the well known halo nucleus,
$^{11}$Be, as a test case. The results show an important predictive power that opens a door 
to the understanding of other lesser known halo nuclei. In order to
show the potential usefulness of the method, it is applied to analyze
the structure of $^{19}$C. 
\end{abstract}

\pacs{21.10.Jx, 21.60.-n 24.10.Eq, 27.20.+n}
\maketitle

\section{\label{intro} Introduction}
One of the main topics in Nuclear Physics during the last years is the
study of nuclei far off the stability line. For these nuclei the ratio
of protons to neutrons is quite different from the usual ratios for
stable nuclei. Because of that, they are known as exotic nuclei. New
physics is expected for these nuclei since, for instance, new close
(sub-)shells could appear affecting to both the structure of the
nucleus and its behavior when participating in nuclear
reactions. Among the observed exotic nuclei, special interest has been
devoted to halo nuclei. These are weakly-bound systems composed by one or two
weakly-bound nucleons orbiting a relatively compact \textit{core}. In
the extreme weak-coupling limit, it is commonly assumed that the
properties of this composite system are mainly determined by the
degree of freedom of the weakly-bound nucleon(s), commonly referred to
as halo. 

In this work we concentrate in two-body halo systems. In
the simplest approach, the halo particle is assumed to move
in a spherical mean-field potential generated by the remaining
nucleons, leading to a description of the levels of the composite
system in terms of single-particle orbitals.  This simple picture is
at the basis of many few-body reaction formalisms used in the analysis
of reactions  induced by halo and other weakly-bound nuclei, such as
the Continuum-Discretized Coupled-Channels (CDCC) method \cite{CDCC}, 
the adiabatic approximation \cite{Ban00,Tos98}, the Faddeev/AGS equations  \cite{faddeev60,Alt}, 
and a variety of  semi-classical approaches \cite{Typ94,Esb96,Kid94,Typ01,Cap04,Gar06}.  
Furthermore, in many of these applications, the valence-\textit{core}
potential is approximated by a simple phenomenological potential, with
the parameters adjusted to reproduce the low-lying spectrum of the
composite nucleus.  
This simple  picture can be  improved including some excited states of
the \textit{core} nucleus. These configurations are naturally included
in microscopic approaches, such as the shell-model and {\it ab-initio} approaches. Within an
effective two-body Hamiltonian, these
\textit{core}-excited components are usually included assuming a collective model for
the \textit{core} nucleus (e.g.~rotor or vibrator) giving rise to the
so-called particle-rotor  \cite{BM} or the particle-vibrator
\cite{Vin95,Gor04} models. In these models, in addition to the central
potential, the valence-\textit{core} interaction contains some
non-central term, which is responsible for the coupling between
different \textit{core} states and gives rise to \textit{core}-excited
admixtures in the states of the composite system. In practice, this is
usually done adding a transition potential with some phenomenological
radial shape and a strength depending on some collective
parameter. The parameters for the central and transition potentials
are usually determined from the known properties of the composite
system and, consequently, require some {\it a priori} knowledge of the
properties of the system, such as the energy excitation and spin-parity
assignment of the low-lying states. This restricts the predictive power of these
models. Moreover, since the rotor and vibrator models are expected to
be limiting cases, it is not guaranteed their accuracy in specific
cases.  

To overcome these limitations, it would be desirable to construct a
particle-plus-\textit{core} model starting from more fundamental
principles. 
Structure models based on microscopic many-body calculations are also potentially useful methods but their applications to halo nuclei with a deformed \textit{core} are still limited~\cite{Des13}. 
Alternatively,  when some properties of
the \textit{core} nucleus are known (rms radius, excitation energies,
etc) one can make use of a semi-microscopic picture, and  
construct the interaction between  
the valence particle and the \textit{core} nucleus by folding  a suitable nucleon-nucleon
effective interaction  with an appropriate \textit{core} density, following the same scheme used in the 
calculation of folding potentials for elastic and inelastic scattering \cite{Sat79}. 

It is our goal in this work to apply this idea to calculate the
energies and wavefunctions of the states of one-neutron halo
nuclei. The aforementioned folding method is applied to generate the
spin-independent part of the particle-\textit{core} interaction.  Both
central and transition potentials are calculated within this scheme,
making use of the appropriate monopole and transition densities. In the
calculations presented in this work, the nucleon-nucleon interaction
of Jeukenne, Lejeune and Mahaux  \cite{JLM} is used as effective
interaction, whereas the \textit{core} densities are calculated here 
using the antisymmetrized molecular dynamics (AMD) method
\cite{Kan95a,*Kan95b}.  A  phenomenological spin-orbit part, with
standard parameters, is also added to the model Hamiltonian. It should be noted that the method can be equally applied with any other appropriate nucleon-nucleon interaction and/or different method to extract the \textit{core} transition densities.




The paper is structured as follows. In Sec.~\ref{sec:weakcoup}
the particle-\textit{core} Hamiltonian is defined and the method
used to obtain the solutions (eigenfunctions and eigenenergies) of this
Hamiltonian is explained. In Sec.~\ref{sec:calc}, the method is
applied to $^{11}$Be and $^{19}$C. Finally,
Sec.~\ref{sec:summary} is devoted to discuss  and summarize the
main conclusions of this work.

\section{\label{sec:weakcoup} \textit{Core} excitations in the structure of two-body halo nuclei}

\subsection{Particle-\textit{core} model}
We consider a composite nucleus, described as a two-body system,
comprising a weakly-bound nucleon coupled to a \textit{core}. In the
weak-coupling limit, the Hamiltonian of the system can be written as 
\begin{equation}
\label{fullham}
 {\cal H}= h_{core}(\xi)+T(\vec{r})+V_{vc}(\vec{r},\xi),
\end{equation}
where $T(\vec{r})$ is the kinetic energy operator for the relative
motion between the valence and the \textit{core}, $h_{core}(\xi)$ is
the  Hamiltonian of the \textit{core}  and $V_{vc}(\vec{r},\xi)$ is
the effective valence-\textit{core} interaction. The variable  $\xi$
denote the internal coordinates of the \textit{core}. The dependence
of $V_{vc}(\vec{r},\xi)$ on these coordinates account for
\textit{core}-excitation effects.

The eigenfunctions of this Hamiltonian, for a given energy $\varepsilon$,
would be characterized by the total angular momentum $\vec{J}$,
resulting from the coupling of the angular momentum $\vec{j}$ of the
valence particle to the \textit{core} angular momentum
$\vec{I}$. These functions can be generically expressed as 
\begin{equation}
\label{wfx}
\Psi_{\varepsilon; J M }(\vec{r},\xi) 
 =  \sum_{\alpha} R^J_{\varepsilon,\alpha}(r) 
\left[  {\cal Y}_{(\ell s) j}(\hat{r}) \otimes \phi_{I}(\xi) \right]_{JM} ,
\end{equation}
where $\phi_{I}(\xi)$ denotes the \textit{core} eigenstates,
$\vec{\ell}$ is the orbital angular momentum between the valence particle and
\textit{core}, which couples to the spin of the valence particle ($\vec{s}$) to
give the particle total angular momentum $\vec{j}$. The label ${\alpha}$ denotes the
set of quantum numbers $\{\ell,s,j\}$. The radial functions
$R^J_{\varepsilon,\alpha}(r)$ can be determined in several ways. A
common procedure is to insert the expansion (\ref{wfx}) into the
Schr\"odinger equation, giving rise to a set of coupled differential
equations for the radial functions $R^J_{\varepsilon,\alpha}(r)$ (see
e.g.~\cite{BM}).  

In this work, instead of this coupled-channels method, we use the
so-called pseudo-state (PS) method. This method consists in
diagonalizing the Hamiltonian in a finite basis of square integrable
functions. This basis is chosen of the form 
\begin{equation}
\phi^{THO}_{n,\alpha,J,M}(\vec{r},\xi)=R^{THO}_{n,\ell}(r) \left[
  {\cal Y}_{(\ell s) j}(\hat{r}) \otimes \phi_{I}(\xi)
  \right]_{JM}\quad 
\label{fTHO}
\end{equation}
where $R^{THO}_{n,\ell}(r)$ are a set of square-integrable radial
functions. For the latter, we use the analytic Transformed Harmonic
Oscillator (THO) basis, which is obtained by applying a local scale
transformation (LST) to the spherical HO basis as 
\begin{equation}
\label{eq:tho}
R^{THO}_{n, \ell}(r)= \sqrt{\frac{ds}{dr}} R^{HO} _{n, \ell}[s(r)],
\end{equation}
where $R ^{HO}_{n, \ell}(s)$ (with $n=1,2,\ldots$) is the radial part
of the HO functions and $s(r)$ defines the LST. For the latter we use
the analytical prescription of Karataglidis \etal \cite{Amos}  
\begin{equation}
\label{lst}
s(r)  = \frac{1}{\sqrt{2} b} \left[  
 {\left(  \frac{1}{r} \right)^m  +  \left( \frac{1}{\gamma\sqrt{r}}
 \right)^m } \right]^{-\frac{1}{m}}\ , 
\end{equation}
that depends on the parameters $m$, $\gamma$ and the oscillator length
$b$. In Ref.~\cite{Amos}, it was stated that the LST depends very weakly on $m$ and they suggested the value $m=4$. In this work we adopt this value, so the only active parameters in the LST are $b$ and $\gamma$. The ratio $\gamma/b$ determines the range of the basis functions
and the density of eigenstates as a function of the excitation energy:
as $\gamma$ decreases, the basis functions explore larger distances
and the corresponding eigenvalues concentrate at lower
excitation energies. Further details are given in
Refs.~\cite{Lay10,Lay12}. 

The eigenstates of the Hamiltonian (\ref{fullham})  will be expressed
as an expansion in the THO basis, $\Phi^{(N)}_{i,J}=\sum_{n=1}^{N}
\sum_\alpha C^{i}_{n,\alpha,J}\phi^{THO}_{n,\alpha,J}$, where $N$ is
the number of radial functions retained in the THO basis, $i$ is an
index labeling the eigenstates for a given $J$, and
$C^{i}_{n,\alpha,J}$ are the expansion coefficients in the truncated
basis.  
The negative eigenvalues of the Hamiltonian (\ref{fullham}) are
identified with the energies of the bound states, whereas the positive
ones provide a discrete representation of the continuum
spectrum. For small values of $N$, some of the positive-energy
eigenvalues become stable with respect to small changes of $N$ or of
some non-linear parameter of the basis (e.g.~$\gamma$). These
stabilized energies are identified with the resonances of the system~\cite{Haz70,Lay12proc}.

\subsection{Matrix elements of the Hamiltonian in the PS basis}
The diagonalization of the Hamiltonian (\ref{fullham}) requires the
evaluation of the matrix elements of  the potential $V_{vc}$  
 in the PS basis, denoted in ket form as $| n (\ell s)j I; J
 \rangle$. For this purpose, it is convenient to separate the angular
 part by performing a multipole expansion of this interaction, i.e.: 
\begin{equation}
 V_{vc}(\vec{r},\xi)= \sum_{\lambda \mu} V_{\lambda \mu}(r,\xi)Y^*_{\lambda \mu}(\hat{r}) = \sum_\lambda V_{\lambda}(r,\xi) \cdot Y_{\lambda}(\hat{r}).
\label{vtrans}
\end{equation}
Then, following the convention for matrix elements used in Brink and
Satchler \cite{BS}, we obtain for each $\lambda$: 
\begin{align}
\label{redpot}
& \langle n (\ell s)j I; J  || V_{\lambda}(r,\xi) \cdot Y_{\lambda}(\hat{r})
|| n' (\ell' s)j' I'; J  \rangle   =     
\nonumber & \\
& \quad \quad \quad   \times (-1)^{(j'+I+J)} (2I+1)^{1/2}  \langle n \ell I
\| V_{\lambda}(r,\xi) \| n' \ell' I' \rangle  \nonumber  \\    
& \quad \quad \quad  \times (2j+1)^{1/2}   \left\lbrace
\begin{array}{ccc} j & j' & \lambda \\ I'& I & J  \end{array}
\right\rbrace 
 \langle (\ell s)j \| Y_{\lambda } \| (\ell's')j' \rangle ,
\end{align}
where
\begin{align}
\langle n \ell I \| V_{\lambda}(r,\xi) \| n' \ell' I' \rangle & = 
   \int  dr ~r^2 R^{J~*}_{n\ell}(r) R^{J}_{n'\ell'}(r)  \nonumber \\
  &\times \langle I \| V_{\lambda} (r,\xi) \|  I' \rangle , 
\end{align}
which contains the dependence of the assumed model for the
\textit{core}. 
For example, in the rotor model,
these matrix elements read (see e.g.~\cite{BM,Tam65,Tho09,Lay12}) 
%
\begin{equation}
\langle I || V_{\lambda}(r,\xi) || I' \rangle= V_{\lambda}(r)
(-1)^{I-I'} \langle I K \lambda 0 | I' K \rangle , 
\end{equation}
where $K$ is the projection of the angular momentum on the \textit{core} symmetry
axis (usually $K=0$ for the lowest energy levels in even-even systems). Each $V_{\lambda}(r)$ reads:
\begin{eqnarray}
V_{\lambda}(r)= \int d\Omega~~ V(r-R(\Omega))  Y_{\lambda 0}(\theta,0) , \\
 R(\Omega)=R_{0}+\sum_{\lambda \geq 2} \delta_{\lambda}Y_{\lambda 0} (\theta,0),
\label{vdef}
\end{eqnarray}
where usually we use a typical Woods-Saxon form for $V(r-R(\Omega))$
and only consider $\lambda=0,2$, being $\delta_{2}=\beta_2R_0$ the deformation
length of the \textit{core}. 



\subsection{Folding model for the valence-\textit{core} interaction}

In this work, we propose a simple semi-microscopic prescription, in which
the  $V_{vc}(\vec{r},\xi)$ interaction is calculated by means of a
folding procedure, convoluting an effective in-medium NN interaction
with microscopic transition densities of the \textit{core} nucleus,
i.e. 
\begin{equation}
V_{vc}(\vec{r},\xi)= \int  d\vec{r'} \rho(\vec{r'},\xi) v_{nn} (\vec{r}-\vec{r'}).
\label{fold}
\end{equation}
where $v_{nn}$ is the effective NN interaction and
$\rho(\vec{r'},\xi)$ the density operator, defined as usual as 
\begin{equation}
 \rho(\vec{r},\xi) =  \sum_{i=1}^{A} \delta(\vecr -\vecr_{i}) . 
\label{densop}
\end{equation}
This is conveniently expanded in multipoles as
\begin{equation}
 \rho(\vec{r'},\xi)= \sum_{\lambda \mu} \rho_{\lambda
 \mu}(r',\xi)Y^*_{\lambda \mu}(\hat{r'}). 
\label{rhotrans}
\end{equation}
 Note that, in the spherical case, $\rho(\vec{r'})=\rho(r')$, and
 $V_{vc}(r)$ becomes a central potential, so it contains only the
 $\lambda=0$ term.  In a
 more general case, as we consider here, $\rho(\vec{r'},\xi)$ contains
 also non-central terms that will give rise to transition terms
 with $\lambda > 0$ in the valence-\textit{core} potential.  

According to Eq.~(\ref{redpot}), one requires the reduced matrix
elements of the $V_{vc}$ interaction between different \textit{core}
states. In the folding scheme, these will be related to the
matrix elements of the density operator between different \textit{core} states, i.e. 
\begin{align}
\langle I \nu | \rho(\vec{r},\xi) | I' \nu' \rangle &= 
   \langle \phi_{I \nu}(\xi) | \sum_{i=1}^{A} \delta(\vecr -\vecr_{i})
   |  \phi_{I' \nu'}(\xi) \rangle \nonumber \\ 
   & = \sum_{\lambda,\mu} \langle I' \nu' \lambda \mu | I \nu \rangle
   \rho_{\lambda,  I' \rightarrow I}(r) Y^{*}_{\lambda \mu}(\hat{r})  
\label{densred}
\end{align}
where $\rho_{\lambda, I' \rightarrow I}(r)$ correspond to the reduced matrix elements
\begin{equation}
 \dent{\lambda}{ I'}{I}(r) \equiv \langle I || \rho_{\lambda} ||  I'\rangle \, .
\label{rhotransdef}
\end{equation}

Our convention for reduced matrix elements is that of Brink and
Satchler \cite{BS} so that the reversed densities are related as 
$
\sqrt{2I'+1}  \langle I' || \rho_{\lambda} || I \rangle \ =\sqrt{2I+1}
\langle I || \rho_{\lambda} || I' \rangle  . 
$

The density operator can be analogously defined for protons and
neutrons ($\rho^{(p)}$ and $\rho^{(n)}$), in which case the sum in
Eq.~(\ref{densop}) runs over protons or neutrons, respectively. The
corresponding monopole transition densities are normalized as 
\begin{align}
 \int d\vec{r}~ \rho_{0,I \rightarrow I}^{(p)}(r) Y_{0 0}(\hat{r}) &= Z, \\
 \int d\vec{r}~ \rho_{0,I \rightarrow I}^{(n)}(r) Y_{0 0}(\hat{r}) &= N.
\label{normdens}
\end{align}
For the proton case, the multipole terms are constrained by the
electric transition probabilities, i.e.: 
\begin{eqnarray}
{\cal B}(E\lambda,  I'\rightarrow I) = \frac{2I+1}{2 I'+1} e^{2} \left |
\int  dr ~ r^{\lambda+2} \dent{\lambda}{ I'}{I}^{(p)}(r)  \right |^2 . 
\end{eqnarray}   
The \textit{core} transition densities can be obtained with different methods. 
In this work, these densities are obtained from
antisymmetrized molecular dynamics (AMD) \cite{Kan99,Kan13a,*Kan13b}
calculations for the \textit{core} nucleus. This method is a microscopic structure model based on effective nuclear interactions in which the antisymmetrization between nucleons is fully taken into account. AMD wavefunctions are formed from Slater determinants of single-nucleon Gaussian wavefunctions.
Namely, many-body wavefunctions are treated without assuming existence of any specific clusters in the method. Nevertheless, the AMD model space covers a variety of cluster structures and it can describe those of neutron-rich nuclei. Actually, the method has been proved to be very useful to understand the level structure and deformation of Be and B isotopes~\cite{Kan95a,*Kan95b}. In the application to Be isotopes, it was shown that the structure of low lying states can be described as two alpha clusters and remaining neutrons around the two alphas as proposed by Von Oertzen~\cite{Oer96}.

Following \cite{Sat79}, the central part of the effective
nucleon-nucleon interaction ($v_{nn}$) is decomposed in terms of the
total spin ($S$) and isospin ($T$) of the colliding pair but, for
simplicity, only the $S=0$ terms are considered, 
\begin{equation}
 v_{nn} (s) =  v_{00} (s) +v_{01} (s)\vec{\tau'}\cdot\vec{\tau} \, ,
\end{equation} 
where  $v_{ST}$ are the expansion  terms and  $\tau$ is the isospin
operator. Attending to the isospin dependence, the $v_{00}$ and
$v_{01}$ terms are  called, respectively, isoscalar and isovector
parts. The radial forms $v_{0T}(s)$ are taken from the work of
Jeukenne-Lejeune-Mahaux (JLM) \cite{JLM} 
\begin{equation}
 v_{0T}(s,\rho,E)=\lambda_{v}V_T(\rho,E)(t_{v}\sqrt{\pi})^{-3} \exp(-s^2/t_{v}^{2}),
\end{equation}
where the strength of the potential, $V_T$, depends on the density
$\rho$, and the nucleon-nucleon relative energy $E$. In this case, for
simplicity, we choose $E=0$. On the other hand, normalizations
factors, $\lambda_{v}$, and the effective range of the Gaussian form
factor, $t_{v}$, are adjustable parameters with typical values between
$0.8$ and $1.2$ for $\lambda_{v}$, and between $1.2$ and $1.4$ for
$t_{v}$. This interaction has been found to reproduce satisfactorily the
elastic and inelastic experimental cross sections in the intermediate
energy region for light  nuclei~\cite{Tak05,Tak08}. 

In order to evaluate Eq.~\eqref{fold} we also expand the
interaction in multipoles as we did for the density: 
\begin{equation}
  v_{0T}(|\vec{r}-\vec{r'}|,\rho,E)= \sum_{\ell}
  v_{0T}^{(\ell)}(r,r')Y_{\ell}(\hat{r})\cdot Y_{\ell}(\hat{r'}) \, . 
\end{equation}

In the test cases considered in this work, the valence particle is a
neutron, in which case  the resulting potential can be expressed in
terms of the corresponding proton and transition densities
as~\cite{Tak05}: 
%
\begin{eqnarray}
\nonumber \langle I || V_{\lambda}(r,\vec{\xi}) || I' \rangle &=  &
 \int dr' ~r'^2  \left \lbrace v^{(\lambda)}_{00}  (r,r')  \right. \\  
& \times & 
 \left[  \rho^{(n)}_{\lambda,I' \rightarrow I}(r) +
 \rho^{(p)}_{\lambda,I' \rightarrow I}(r) \right]  \nonumber \\ 
 &  + &  v^{(\lambda)}_{01} (r,r')  \nonumber \\ 
 &\times & \left. \left[ \dent{\lambda}{I'}{I}^{(n)}(r') -
 \dent{\lambda}{I'}{I}^{(p)}(r') \right]  \right\rbrace  .  
\end{eqnarray}
Note that, if the valence particle is a proton, the signs in the isovector part are changed. 
\section{\label{sec:calc}  Application to halo nuclei}

\subsection{Structure of $^{11}$Be}

As a test example of the formalism presented in the preceding
section, we first consider the well known one-neutron halo nucleus
$^{11}$Be.  
Although the low-lying spectrum of this nucleus is reasonably
described in terms of single-particle configurations, it is known
that these states contain significant admixtures of
\textit{core}-excited components. To account for these components,
within a particle-plus-\textit{core} picture, several models have been
used, such as  the particle-vibrator (PVM) model \cite{Vin95,Gor04}
and the particle-rotor (PRM) model \cite{Tar04,Nun96}. Pairing effects
have also been treated approximately within the quasi-particle rotor
(QPRM) and quasi-particle vibrator models (QPVM) \cite{Tar06}. 

Here, we compare the semi-microscopic approach proposed in this work with
the  particle-rotor model (PRM)  developed by Nunes \etal
\cite{Nun96a,Nun96} (model Be12-b). This model accounts well for the
energies of the bound states and low-lying resonances $3/2^+_1$,
$5/2^+_1$ and $3/2^-_1$ and has been previously employed to illustrate the
use of the PS-THO basis described above \cite{Lay12}. 

The required transition densities in $^{10}$Be are obtained from the
antisymmetrized molecular dynamics  (AMD) calculation of
Ref.~\cite{Kan99}. The AMD model is able to reproduce 
$E2$ transition probabilities between the different $^{10}$Be energy levels 
with a substantial improvement with respect to the shell model
calculations. In addition, the central and transition potentials
calculated with these densities and the JLM potential are able to
reproduce the $p$+$^{10}$Be inelastic cross sections at intermediate
energies \cite{Tak08} within the DWBA framework. This
particle-\textit{core} folding potential based in AMD transition
densities will be referred to as P-AMD.

\begin{figure}
\centering
\epsfig{file=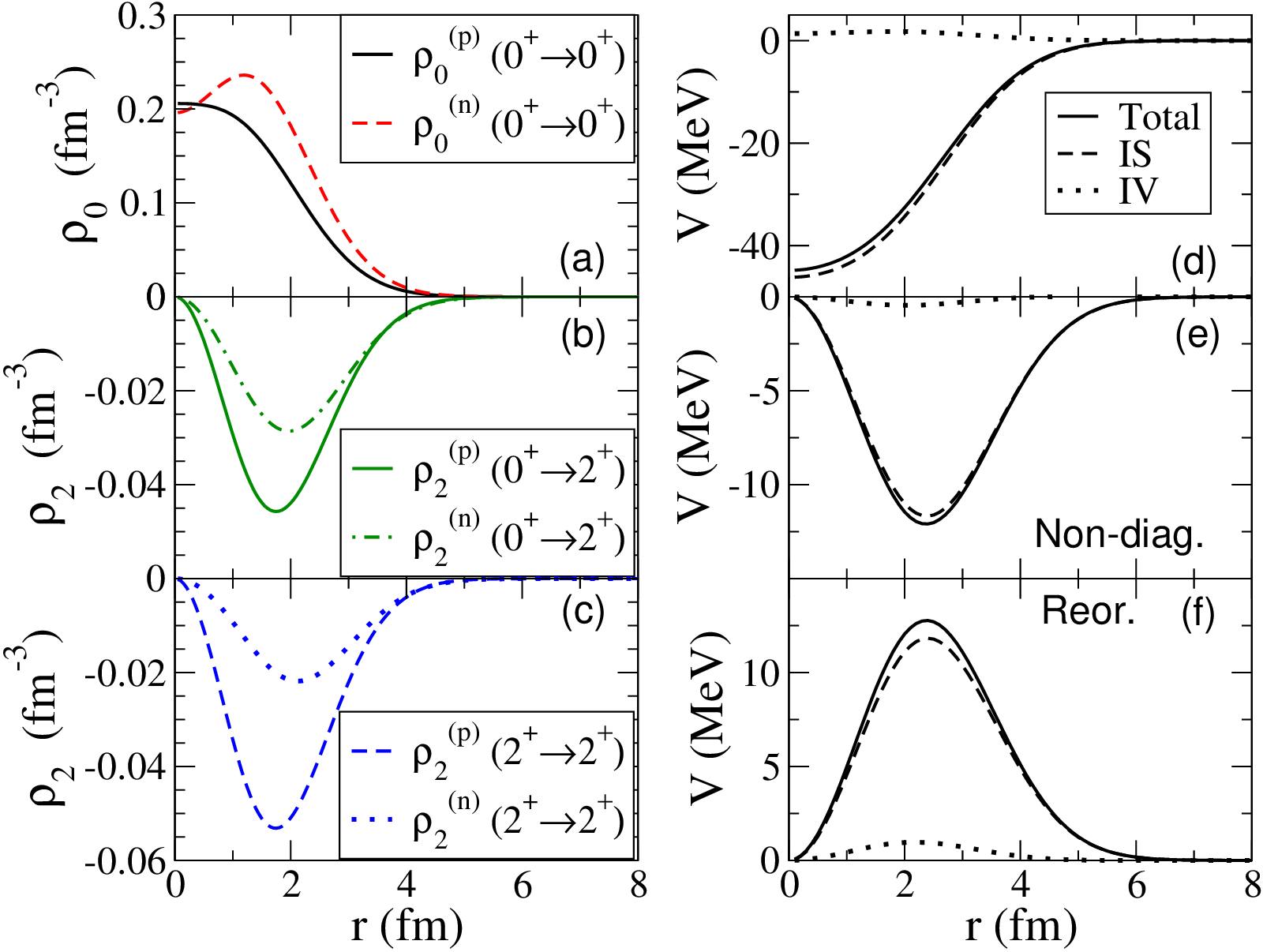,width=\columnwidth}
 \caption{\label{fig:densbe10}  (Color online) Neutron and proton transition densities
 for the two states considered in $^{10}$Be, the 0$^+$ ground state and the 2$^+$
 first excited state. The  central densities,
 $\lambda=0$, are shown in the upper-left panel normalized according
 to  \eqref{normdens} and the quadrupole transition densities, in the
 lower-left panels. The corresponding transition potentials are shown
 in the right panels, with the isoscalar (IS) and isovector (IV) contributions   indicated separately.}  
\end{figure}


In our calculation, we will only consider valence configurations with
 $\ell \leq 2$ and the two lowest lying states in $^{10}$Be, the 0$^+$ ground
 state and the 2$^+$ first excited state. All possible transition
 densities between these \textit{core} states are plotted in
 Fig.~\ref{fig:densbe10}. Convoluting the JLM interaction
 ($\lambda_{v}=1.0$ and $t_{v}=1.2$) with these densities, 
 the potentials shown in
 Fig.~\ref{fig:densbe10} (right panels) are obtained. The
 $\dent{0}{2^+}{2^+}$ density and the 
 corresponding $\langle 2^{+} \| V_{0}\|2^{+} \rangle$ potential are
 included in the calculations although they are not plotted  in
 Fig. \ref{fig:densbe10}. In the rotor model this
 potential coincides with the central $\langle 0^{+} \| V_{0}\|0^{+}
 \rangle$ one.  In our P-AMP model both potentials are almost
 identical as so do the
 corresponding densities, confirming that the rotor assumptions are
 satisfied for the $^{10}$Be \textit{core}. 
 The potentials have been calculated with the code  {\sc MINC} by
 M.~Takashina~\cite{MINC}. 

\begin{figure}
\centering
\epsfig{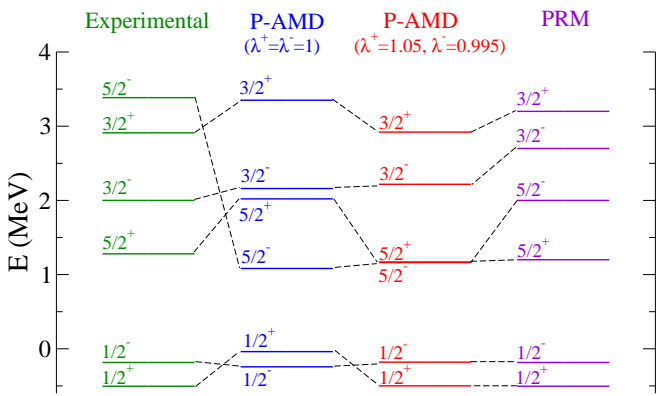}
\vspace{0.50cm}
\caption{ (Color online) Experimental and calculated energy levels of $^{11}$Be. Starting from the left, the second column is the P-AMD calculation without any renormalization. The third column is the P-AMD spectrum obtained with the indicated renormalization constants for positive ($\lambda_{+}$) and negative ($\lambda_{-}$) parity levels. The last column corresponds to the PRM calculation.  Experimental values are from \cite{Fuk04,Hir05}.
 }
\label{fig:niveles11be}
\end{figure}

In addition to the central term, the $n$-$^{10}$Be interaction will
contain spin-dependent parts. For simplicity, we consider only the
spin-orbit term, for which we adopt the phenomenological
parameterization of the potential Be12-b with a standard strength
$V_{so}=6$~MeV.  


The calculated spectrum is shown in 
Fig.~\ref{fig:niveles11be} (second column). Resonant energies are identified with
stabilized eigenvalues with respect to variations in the number of
states included in the THO basis \cite{Lay12}.  
The P-AMD calculation succeeds to produce two weakly bound states
($1/2^+$ and $1/2^-$), in agreement with experiment,  but with the
wrong ordering. Several low-lying resonances ($5/2^\pm$ and
$3/2^\pm$), are also predicted.  
The inversion of the 1/2$^{+}$ and 1/2$^{-}$ levels has been ascribed
to a combined effect of the \textit{core} deformation, Pauli
blocking and pairing effects \cite{Gor04}.  Pairing effects are completely ignored in our treatment whereas Pauli blocking is only considered approximately (see discussion below) so we cannot expect an accurate
description of the experimental spectrum. To account in an effective way for these effects
a slight renormalization of the folding potential is allowed. In order
to reproduce the experimental ordering of the mentioned states, the renormalization factors
need to be different for positive  ($\lambda_{+}$) and
negative  ($\lambda_{-}$) parity states: 
$\lambda_{+}=1.054$ and $\lambda_{-}=0.995$. The new spectrum is also
shown in Fig.~\ref{fig:niveles11be} (third column). The position of the resonances
$3/2^-$,  $5/2^+$ and $3/2^+$ are now reasonably well reproduced. Only
the $5/2^-$ resonance is not well reproduced by the model. Considering
that the only adjustable parameters are $\lambda_{\pm}$, and that the
required normalizations are of the order of 5\%, we can conclude that
the overall agreement is fairly good.  It is worth mentioning that the
PRM model requires also a weaker strength for negative
parity states to obtain the inversion. This fact can be related with the
Pauli exclusion principle. The antisymmetrization of the wavefunctions
should add an extra repulsion to $p_{1/2}$ configurations, repulsion
that was added phenomenologically by reducing the strength of the
negative parity potentials~\cite{Nun95}. 

It is also worth noting that, in addition to the levels shown in
Fig.~\ref{fig:niveles11be}, some other deeply bound eigenvalues are
obtained in the diagonalization. These are identified with Pauli forbidden states and are
therefore removed.  These states come from the $1s_{1/2}$, $1p_{3/2}$
and $1p_{1/2}$ orbitals in the spherical basis, which are already
occupied in the $^{10}$Be nucleus according to our simple model in
which exchange and pairing effects are ignored.  By construction, the states obtained here are orthogonal to those removed, what should account for the Pauli principle. The forbidden states are therefore an admixture of different valence+\textit{core} configurations. Note that, with this procedure, the part of the single-particle strengths of the $1s_{1/2}$, $1p_{3/2}$ and $1p_{1/2}$ orbitals will appear embedded among the retained valence+\textit{core} states. Alternatively, Pauli principle could be applied by removing these spherical valence configurations completely, as done for example in Refs.~\cite{Kuk86,Mio07}. Both methods are indeed approximate. Since in the cases treated in this work the \textit{core} is deformed, we follow the former approach. This election slightly affects the energies and spectroscopic factors obtained.



In Fig.~\ref{fig:niveles11be}, we can see that most of the low lying
structures in $^{11}$Be can be understood within the P-AMD model. The
spectra provided by P-AMD and PRM models \cite{Lay12} are
compatible. P-AMD gives excitation 
energies for the positive parity resonances 5/2$^{+}$, and 3/2$^{+}$ in better agreement with experiment,
although a major part of this effect is related to the spin-orbit
term. P-AMD also improves the excitation energy for the negative
parity resonance 3/2$^{-}$.   

In addition to the energies of the bound and resonant states, we have
compared the weights of the various relevant components (channels) for
the different models considered here. Within our assumed two-body
model, these weights can be regarded as spectroscopic factors. 
In Table \ref{Tab:sf} these spectroscopic factors are shown for the
positive parity states in $^{11}$Be calculated for: the  P-AMD model,
the particle-rotor model (PRM), and a shell model calculation with the
WBT interaction from Warburton and Brown \cite{WBT}. The latter was
performed with the code {\sc oxbash} \cite{OXB}. The three
calculations give slightly different but compatible spectroscopic factors. 
In particular, these models agree in the dominance of the $^{10}$Be(0$^+$) component for the ground state and $5/2^+$ resonance, as well as in the dominant $^{10}$Be(2$^+$) contribution in the $3/2^+$ resonance.

\begin{figure}
\centering
\epsfig{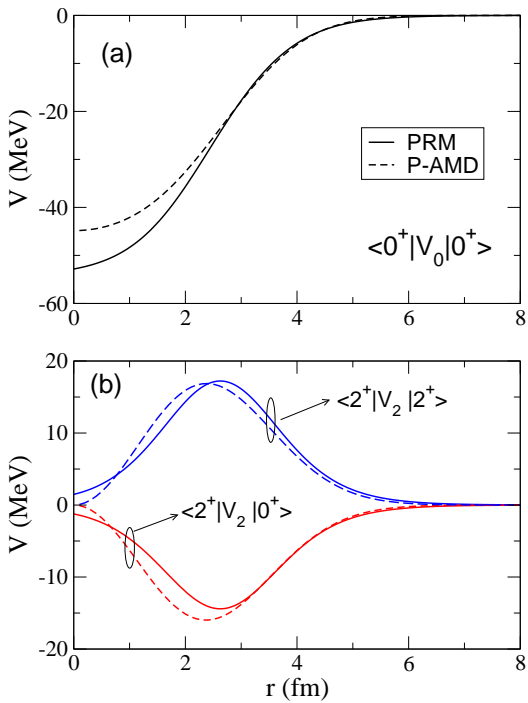}
\caption{ (Color online) Potentials obtained for the $n$-$^{10}$Be system
 with the PRM model and with the P-AMD model. See text for details.}
\label{fig:potjlmvsprm}
\end{figure}

The agreement between P-AMD and PRM values is not
unexpected in view of the similarity of the corresponding transition
potentials. These are shown in Fig.~\ref{fig:potjlmvsprm}. The main difference is
that the P-AMD densities yield a larger deformation of the
\textit{core}. In the PRM this deformation is
a parameter that was estimated from the experimental quadrupole moment of the
\textit{core}, and
corrected by the charge to mass deformation ratio given by shell model
calculations~\cite{Nun96a}. On the other hand, a
deformation parameter can be inferred from AMD densities comparing
the transition density with the derivative of the central density. The
relation between these two magnitudes, assuming that the rotor model
is a good approximation, is
\be
\dent{\lambda}{I'}{I}(r) \approx \langle I \|\hat{\delta}_{\lambda} \| I' \rangle \frac{d\rho_0}{dr} ,
\ee
where $\langle I \|\hat{\delta}_{\lambda} \| I' \rangle$ is the reduced matrix element of the deformation length operator for a given transition.
Multiplying in both sides by $r^2$ and integrating in $r$ one obtains
\begin{equation}
\langle I \|\hat{\delta}_{\lambda} \| I' \rangle =
\frac{1}{\lambda +2}\frac{\int \dent{\lambda}{I'}{I}(r) r^{\lambda+2} dr}{\int \rho_{0}(r) r^{\lambda+1} dr} .
\label{deldens}
\end{equation}
Using the microscopic AMD densities employed in this work, one gets
$ \langle 0 \|\hat{\delta}_{2} \| 2 \rangle_\mathrm{AMD}= 1.90~\mathrm{fm}$. Interestingly, this value is very close to that obtained by Iwasaki \textit{et al} from a DWBA analysis of 10Be(p,p') inelastic data, $1.80\pm0.25$~fm~\cite{Iwa00a}. 
On the other hand, in the rotor model, these matrix elements are related to the intrinsic deformation as
\be
\langle I \|\hat{\delta}_{\lambda} \| I' \rangle  = \langle I K \lambda 0 | I' K\rangle \beta_{\lambda} R_0 .
\label{delrot}
\ee
Inserting the mean radius and deformation parameter employed in our calculations ($R_0=2.483$~fm, $\beta_2$=0.67) one gets 
$\langle 0 \|\hat{\delta}_{2} \| 2 \rangle_\mathrm{rot}= 1.66~\mathrm{fm}$, which is somewhat smaller than the AMD value 
(these values are also listed in Table \ref{Tab:def} for convenience). 
Consequently, the effective deformation obtained from the AMD densities is larger than that  assumed in the  rotor model used here and this explains the larger mixing of the $^{10}{\rm Be}(2^+)$ component in the $^{11}{\rm Be}$ wavefunctions.

      

\begin{table}
\caption{\label{Tab:sf} Spectroscopic factors for the ground state and low-lying positive energy resonances in  $^{11}$Be, according to the different models considered.}
\begin{center}
\begin{tabular}{cccccc}
\hline
  & Model & $| 0^+ \otimes (\ell s)j \rangle $ &    $ | 2^+ \otimes s_{1/2} \rangle $  &   $ | 2^+ \otimes d_{3/2} \rangle $ & $ | 2^+ \otimes d_{5/2} \rangle $    \\
\hline  
\hline
  $1/2^+$      &  PRM    & 0.857    &    --   &     0.021              &    0.121    \\
               &  P-AMD   & 0.849   &      -- &        0.031  &  0.121   \\
               &  WBT    & 0.762    &      -- &        0.002          & 0.184  \\  
\hline
        $5/2^+$     &   PRM    &  0.702   &  0.177  &  0.009  &  0.112  \\ 
                    &   P-AMD  &  0.674   &  0.189  &  0.014  &  0.124 \\   
                    &   WBT    &  0.682   &   0.177 &  0.009  &  0.095       \\ 
\hline
         $3/2^+$                  &      PRM       & 0.165 &  0.737  & 0.017  & 0.081   \\  
                   &     P-AMD       & 0.316 &   0.565   &  0.031  &  0.089   \\
                          &     WBT     &  0.068   &   0.534   &  0.008   &      0.167  \\
\hline
\end{tabular}
\end{center}
\end{table}

\begin{table}
\caption{\label{Tab:def} Properties of the  $^{10}$Be and $^{18}$C systems, derived from the AMD and rotor calculations. The root mean square radius (rms) is obtained from the corresponding central AMD density.}
\begin{center}
\begin{tabular}{cccccc}
\hline
Nucleus     &  rms  &  $\langle 0 \|\hat{\delta}_{2} \| 2 \rangle_\mathrm{AMD}$   &  $\langle 0 \|\hat{\delta}_{2} \| 2 \rangle_\mathrm{rot}$ \\
            &  (fm) &      (fm)                                                   & (fm)          \\
\hline
\hline
$^{10}$Be   & 2.538  &   1.90                                                      & 1.66 \footnote{Rotor model Be12-b of  Ref.~\cite{Nun96}.}  \\                 
$^{18}$C    & 2.776  &   1.20                                                      & 1.50 \footnote{Rotor model of Ref.~\cite{Tar04}, referred in this work as PRM(1).} \\                         
\hline 
\end{tabular}
\end{center}
\end{table}

From the results presented in this subsection, we can conclude that the
developed P-AMD model gives an overall good description of the bound
states and low-lying resonances in $^{11}$Be with very small
adjustments of the parameters involved in the calculation. These
results encourage us for using this model to make predictions for the
structure of poorer known halo nuclei. As an example, in the next
subsection we apply the model to $^{19}$C.

\subsection{Structure of $^{19}$C}
Once tested the semi-microscopic model with the well known nucleus
 $^{11}$Be, we now consider the halo nucleus $^{19}$C. 
 The properties of this nucleus are not so well known and the experimental data, needed to
 adjust the parameters involved in phenomenological models (as PRM or
 PVM), are scarce and, sometimes, contradictory. Therefore, a
 semi-microscopic model as P-AMD can shed some light on the structure of
 this nucleus.  

 Although the properties of this nucleus, including the low-lying
 spectrum, are not well known, it has been recently the focus of several
 works  \cite{Nak99,Typ01b,Ham07,Sat08,Kar08}. It is known that the
 ground state has spin and parity 1/2$^{+}$ and that the binding
 energy with respect to the $^{18}$C+$n$ threshold is
 $0.58\pm0.09$~MeV \cite{Aud03}. Almost all theoretical calculations
 predict a 1/2$^{+}$ (prolate) and a 3/2$^{+}$ (oblate) almost
 degenerate states \cite{Suz03}. In fact, the ground state spin and parity were not
 confirmed until recently \cite{Sat08}. The deformation for all lighter
 carbon isotopes is known to be prolate. For $^{19}$C, prolate and
 oblate structures seem to be almost degenerate (shape coexistence)
 anticipating a kind of shape phase transition. It can be
 indicative of the presence of a new magic number, $N=16$, for neutron
 rich nuclei \cite{Suz03}.

\begin{figure}
\centering
\epsfig{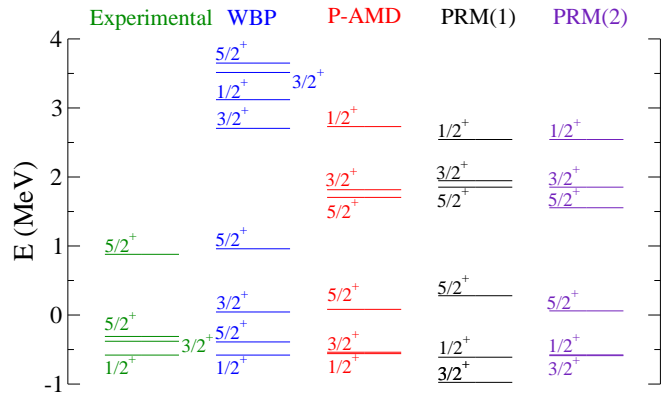}
\caption{(Color online) Spectrum obtained for the $^{19}$C nucleus within the two PRM calculations (PRM(1) and PRM(2)), a  shell model calculation (WBP), and  with the single-folding calculation based on microscopic densities of the \textit{core} (P-AMD) compared with the experimental one \cite{Sat08,Ele05}. }
\label{fig:niveles19c}
\end{figure}

In addition to the $1/2^+$ and $3/2^+$ bound states, the analysis of
the inelastic data of $^{19}$C on protons reported in
Ref.~\cite{Ele05} suggested the existence of another bound excited state.
This was assigned a spin-parity $5/2^+$ with the guidance of shell model
calculations. Later on, in a exclusive breakup experiment of 
$^{19}$C+$p$, a prominent peak was observed in the relative energy
spectrum of the outgoing $^{18}$C and $n$ particles. Using microscopic
DWBA calculations based on shell model densities, this state was
associated with a second $5/2^+$ state predicted by some shell model
calculations. This is the accepted experimental knowledge of the
low-lying $^{19}$C spectrum presented in the left part of
Fig.~\ref{fig:niveles19c}. However, the experimental data are far from being clearly established. For example, a recent knockout
experiment \cite{Kob12} seems to question the existence of a $5/2^+$
bound state. From the
theoretical point of view, the situation is also unclear. The
quasi-particle rotor model (QPRM) of Ref.~\cite{Tar06} gives correctly
the $1/2^+$ ground state, but predicts that the first excited state is
$5/2^+$.  In Ref.~\cite{Kar08}, the  $^{19}$C was studied within a
multi-channel algebraic method based upon a two-state, collective,
model for the $n$+$^{18}$C system. In order to reproduce the triplet of bound states reported
by Elekes \etal \cite{Ele05} as well as the $5/2^+$ resonance suggested by Satou \etal \cite{Sat08} they need 
to introduce some Pauli hindrance of the $1d_{5/2}$ orbit. This introduces a phenomenological parameter in the model, which 
accounts for the amount of Pauli blocking of a given orbital. 
 Clearly, the situation calls for further experimental and theoretical
works.  

\begin{figure}
\centering

\epsfig{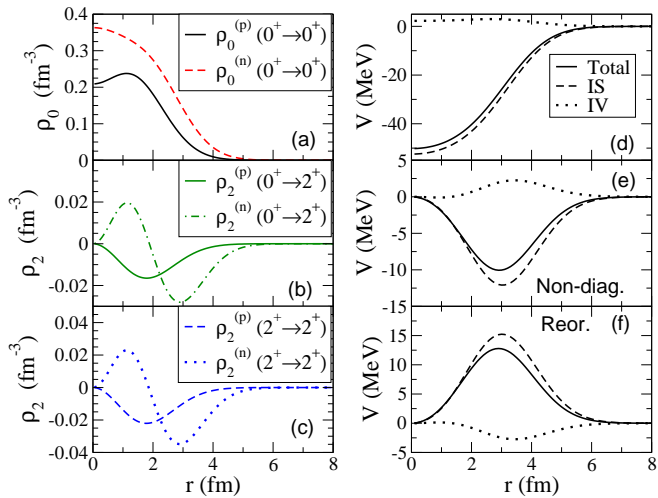}

\caption{ (Color online) Neutron and proton transition densities between the two considered lowest lying states in $^{18}$C,
  the 0$^+$ ground state and the 2$^+$ first excited state. The  central densities, $\lambda=0$, are shown
  in the upper-left panel normalized according to  \eqref{normdens}
  and the quadrupole transition densities, in the lower-left
  panels. The corresponding transition potentials are shown in the
  right panels.} 
\label{fig:densc18}
\end{figure}

With the aim of shedding some light into this problem, we present our
prediction for the structure of $^{19}$C within the semi-microscopic P-AMD
framework. As in the $^{11}$Be case, the neutron+$^{18}$C folding
potential was generated with the JLM nucleon-nucleon interaction, and
the monopole and transition densities calculated with AMD.  These
densities, and the corresponding transition potentials, are shown in  
Fig.~\ref{fig:densc18}. The central  folded potential is supplemented with  a phenomenological spin-orbit term, parameterized in terms of the derivative of a Woods-Saxon shape, with a standard strength $V_{so}=6.5$~MeV. 
The geometry is adjusted to be consistent with the extension of the central part of the folding potential, 
obtaining $R_{so}=3.0$~fm and $a_{so}=0.70$~fm.

\begin{figure}
\centering
\epsfig{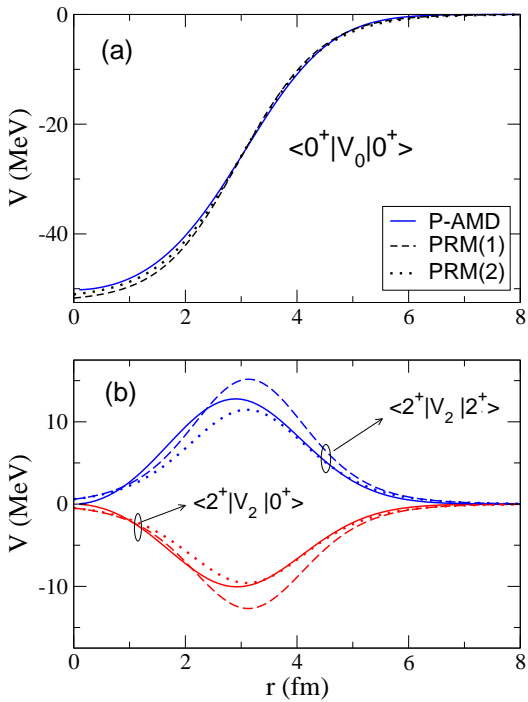}
\caption{ (Color online) Potentials obtained for the $n$-$^{18}$C system with the PRM and P-AMD models.}
\label{fig:potjlmc19}
\end{figure} 
Diagonalizing the Hamiltonian, the spectrum shown under P-AMD in
Fig.~\ref{fig:niveles19c} is obtained. The experimental levels
reported in Refs.~\cite{Sat08,Ele05}, with the spin-parity assignment
suggested in these works, are indicated in the left part of the figure. The
prediction of two PRM calculations, labeled as PRM(1) and PRM(2), are
shown at the right part of the figure. 
PRM(1) is from Tarutina and Hussein~\cite{Tar04} that use a Woods-Saxon
central potential with parameters $V_0=-52.30$~MeV, $V_{so}=6.5$~MeV,
$R_0=3.0$~fm and $a_0=0.65$~fm and a transition potential, obtained by
deforming the central potential with a deformation parameter of
$\beta_2=0.5$~fm. On the other hand, the model PRM(2) is obtained by
adjusting the Wood-Saxon parameters to reproduce the potential provided by the P-AMD model. The
aim of this PRM(2) model is to reproduce P-AMD results with simpler
potentials in a more standard and widespread framework, so that these
results can be more easily reproduced and used for different
purposes. From this adjustment it is obtained $V_0=-51.80$~MeV,
$R_0=3.0$~fm and $a_0=0.70$~fm and assumes a deformation parameter
$\beta_2=0.4$~fm. Also shown in 
Fig.~\ref{fig:niveles19c}, is the spectrum predicted by the shell-model calculation performed with the WBP 
effective interaction of Warburton and Brown \cite{WBT}.
As expected, the PRM(2) model reproduces the P-AMD spectrum quite well. 

The P-AMD model predicts a $1/2^+$ ground state, with a separation
energy of $S_n=-\varepsilon_{gs}=0.582$~MeV, in excellent agreement
with the experimental value. The first excited state is $3/2^+$, also
in agreement with the experimental data, although in our model this
state is almost degenerate with the ground state. No additional bound
states are found in this model, contrary to the suggestions of
\cite{Ele05}. The first resonant state is a $5/2^+_1$, which appears
very close to the threshold. Taking into account the approximations
implied in our model, one cannot rule out that this state is actually
a weakly bound state, as suggested in  \cite{Ele05}. A second
$5/2^+$ resonance is obtained in P-AMD at 
$\varepsilon=1.704$~MeV, but this state does not have a clear
counterpart in the experimental spectrum. No states are found close to
the resonant peak observed by Satou \etal \cite{Sat08} which, again,
could be attributed to the uncertainties of our model. The PRM(1) model predicts also two bound states with spin and parity $1/2^+$ and $3/2^+$, but with the latter being lower in energy. As in the P-AMD model, the second $5/2^{+}$ state appears as a low-lying resonance close to the neutron separation threshold.

 The transition potentials obtained with the P-AMD, PRM(1) and PRM(2) models are compared in 
Fig.~\ref{fig:potjlmc19}. The central potentials are similar in the three models, whereas the quadrupole transition potentials 
($\langle 2^{+} \| V_{2}\|2^{+} \rangle$ and  $\langle 2^{+} \| V_{2}\|0^{+} \rangle$) are larger in the PRM(1) model. 
This can be understood in terms of the corresponding deformation lengths. The values computed with 
 Eqs.~(\ref{deldens}) and (\ref{delrot}) are listed in 
Table \ref{Tab:def}. In this case, the deformation predicted by the AMD model is smaller than the one assumed in the rotor model PRM(1), and this explains the stronger transition potential in the latter case.


\begin{table}
\caption{\label{Tab:sf19c} Spectroscopic factors for the ground state and low-lying positive energy resonances in  $^{19}$C, according to the different models considered in this work.}
\begin{center}
\begin{tabular}{cccccc}
\hline
 & Model & $| 0^+ \otimes (\ell s)j \rangle $ &    $ | 2^+ \otimes s_{1/2} \rangle $  &   $ | 2^+ \otimes d_{3/2} \rangle $ & $ | 2^+ \otimes d_{5/2} \rangle $    \\
\hline  
\hline
$1/2^{+}_{1}$  &  P-AMD    & 0.529    &    --   &  0.035  &   0.436    \\
               &  PRM(1)   & 0.517    &    --   &  0.081  &   0.402    \\  
               &  PRM(2)   & 0.505    &    --   &  0.033  &   0.462    \\  
               &  WBP      & 0.580    &    --   &  0.085  &   0.470    \\

\hline
$3/2^{+}_{1}$  &   P-AMD   & 0.028    &  0.386  &  0.121  &   0.464   \\  
               &   PRM(1)  & 0.043    &  0.348  &  0.150  &   0.459    \\  
               &   PRM(2)  & 0.023    &  0.371  &  0.106  &   0.500    \\  
               &   WBP     & 0.026    &  0.494  &  0.001  &   0.076   \\

\hline
$5/2^{+}_{1}$  &   P-AMD   & 0.276    &  0.721  &  0.000  &   0.003  \\ 
               &   PRM(1)  & 0.285    &  0.716  &  0.000  &   0.003    \\  
               &   PRM(2)  & 0.278    &  0.719  &  0.000  &   0.003    \\  
               &   WBP     & 0.383    &  0.015  &  0.000  &   0.751       \\ 

\hline
$5/2^{+}_{2}$  &   P-AMD   & 0.200    &  0.142  &  0.002  &   0.657   \\  
               &   PRM(1)  & 0.217    &  0.178  &  0.004  &   0.602    \\  
               &   PRM(2)  & 0.207    &  0.100  &  0.002  &   0.690    \\  
               &   WBP     & 0.035    &  0.609  &  0.009  &   0.291  \\

\hline
\end{tabular}
\end{center}
\end{table}

Despite slight disagreements in energy, all studied models give the same spin and parity for the four lowest-lying states. In Table~\ref{Tab:sf19c} the spectroscopic factors provided by the different models for these four levels are presented. Good agreement is found for all the states as in $^{10}$Be. In the case of the two $5/2^+$ states, the ordering predicted by the  PRM and P-AMD models seems to be inverted  with respect to the shell model prediction.  


\begin{figure}
\centering
\epsfig{file=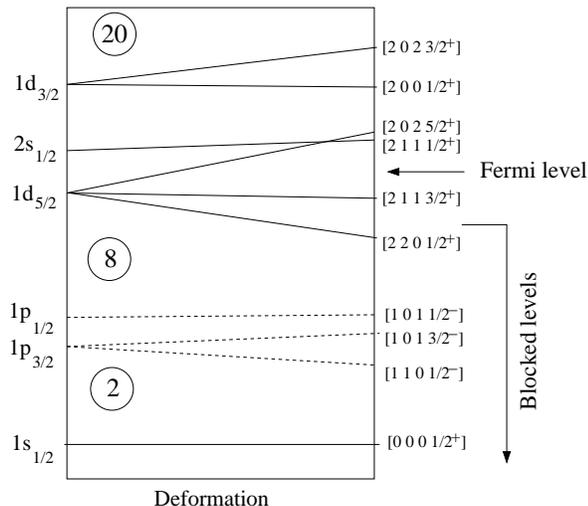,width=0.9\columnwidth}
\caption{Schematic diagram for the lowest Nilsson levels relevant for the calculation of the Pauli forbidden states in our P-AMD model for $^{19}$C. }
\label{fig:Nilsson}
\end{figure}

It is worth mentioning that the original value of the central potential depth in model PRM(1) was $V_0=-42.95$~MeV~\cite{Tar04}. With this choice one obtains, a $1/2^+$ state with the experimental separation energy and a deeply bound $1/2^+$  state, which is considered to be Pauli forbidden. In the calculations presented in this work, we assume that there is an additional 1/2$^{+}$ Pauli forbidden state, and hence the depth of the potential has been increased accordingly in order the third $1/2^+$ eigenvalue has the  experimental separation energy.  The last neutrons in $^{18}$C partially occupy the $2s_{1/2}$ and $1d_{5/2}$ orbits, so that full removal of these spherical configurations may produce misleading results. 
In order to estimate the number of Pauli forbidden states we follow the procedure of Ref.~\cite{Ura11}, which makes use of the strong coupling limit in the simple Nilsson model \cite{BM,Ura11}. Using the asymptotic quantum numbers $[N n_z \Lambda K^\pi]$, corresponding to large prolate deformations, the  relevant Nilsson levels for $^{18}$C (N=12) are: $[0 0 0 \frac{1}{2}^+]$, $[1 1 0 \frac{1}{2}^-]$, $[1 0 1 \frac{3}{2}^-]$, $[1 0 1 \frac{1}{2}^-]$, $[2 2 0 \frac{1}{2}^+]$, $[2 1 1 \frac{3}{2}^+]$, $[2 1 1 \frac{1}{2}^+]$ and $[2 0 2 \frac{5}{2}^+]$ (see scheme in Fig.~\ref{fig:Nilsson}). In this extreme model the occupancy of each level is two. Levels well below the Fermi level are completely occupied and blocked, levels around the Fermi level are only partially occupied and participate in the low-energy excitation of the system. In the $^{18}$C case the five lowest Nilsson levels are fully occupied and blocked. The extra 2 neutrons of the \textit{core} can occupy partially $[2 1 1 \frac{3}{2}^+]$, $[2 1 1 \frac{1}{2}^+]$ and $[2 0 2 \frac{5}{2}^+]$ but do not block completely these levels to the extra neutron in $^{19}$C. In order to calculate the Pauli forbidden states for the extra neutron, $[0 0 0 \frac{1}{2}^+]$, $[1 1 0 \frac{1}{2}^-]$, $[1 0 1 \frac{3}{2}^-]$, $[1 0 1 \frac{1}{2}^-]$ and $[2 2 0 \frac{1}{2}^+]$ levels are blocked. Taking into account that we are only interested in positive parity states in $^{19}$C, and that we are including only the $0^+$ and the first excited state $2^+$ states of the \textit{core}, the negative parity Nilsson orbitals are not important since they will produce negative parity states in $^{19}$C. Consequently, the relevant blocked Nilsson levels are $[0 0 0 \frac{1}{2}^+]$ and $[2 2 0 \frac{1}{2}^+]$. The Nilsson model corresponds to an adiabatic approximation of the particle-\textit{core} model in which the \textit{core} states are assumed to be degenerate in energy. Once the energy of the \textit{core} states is increased the $K$ quantum number is no longer a good quantum number, but one can monitor how the orbits characterized by this quantum number splits into several non-degenerate states, characterized by the total angular momentum $J$ of the system. 
 In particular, one gets
two 1/2$^+$ states ($1/2^+ \otimes 0^+$ from $[0 0 0 \frac{1}{2}^+]$ and  $[2 2 0 \frac{1}{2}^+]$), two 3/2$^+$ states, and two 5/2$^+$ states ($1/2^+ \otimes 2^+$ from $[0 0 0 \frac{1}{2}^+]$ and  $[2 2 0 \frac{1}{2}^+]$) that should be removed. Thus, in our calculations, the third 1/2$^+$ state is considered to be the physical ground state of the $^{19}$C system. The fact that, with this selection of Pauli forbidden states, the ground state energy, its spectroscopic factor and the level scheme, among other observables, are in good agreement with experimental data and shell model calculations as shown in Table~\ref{Tab:sf19c} and Fig.~\ref{fig:niveles19c} supports this procedure of calculating the Pauli forbidden states.

\section{\label{sec:summary} Summary and conclusions}

A semi-microscopic particle-plus-\textit{core} description of one-neutron halo nuclei, considering excitations of the \textit{core}, has been presented.  In this model, the neutron-\textit{core} interaction is constructed by folding an effective nucleon-nucleon interaction with microscopic (monopole and transition) densities of the \textit{core}. For the former, the JLM \cite{JLM} interaction is chosen, whereas the densities are calculated with the Antisymmetrized Molecular Dynamics (AMD) method.  
 The folded potential is supplemented with a phenomenological spin-orbit interaction. 

The model has been applied to the well known halo nucleus  $^{11}$Be ($^{10}$Be+$n$) and to the less known halo nucleus $^{19}$C ($^{18}$C+$n$).  In the  $^{11}$Be case, the model is able to reproduce the experimental spectrum using standard parameters for the JLM interaction in the region of the nucleus. Small normalization factors, different for positive and negative parity states, are required in order to reproduce the separation energies of the two bound states in $^{11}$Be. However, these normalization factors are close to one ($\lambda_{+}=$1.056 and $\lambda_{-}=$0.995 for positive and negative parity states, respectively) suggesting that this semi-microscopic description can have predictive power to study less known nuclei with an important effect of \textit{core} excitations, such as odd-even nuclei in deformed regions. The small difference in the renormalization factors  can be understood considering the effect of Pauli repulsion in the negative parity states due to the nucleons that fill the $1p_{3/2}$ and $1p_{1/2}$ states. The good agreement between the spectroscopic factors from particle-rotor model, shell model and this semi-microscopic description also supports the predictive power of this method. 

Based on these results, AMD densities for $^{18}$C~\cite{Kan13a,*Kan13b} have been used to obtain the spectrum, spectroscopic factors and properties of the $^{19}$C nucleus. Very few experimental data are available for this halo nucleus. Without any renormalization of the coupling potentials, the model reproduces the spin-parity (1/2$^{+}$) and the separation energy (S$_n=0.58$~MeV) of the ground state. It also predicts the existence of a bound 3/2$^{+}$ state, in agreement with shell model calculations and experimental evidences. Two low-lying 5/2$^{+}$ resonances 
have been found. The calculated spectroscopic factors for these two resonances suggest a possible inversion in the energy sequence of these states with respect to shell model. 
Apart from this exception, it should be emphasized that the proposed semi-microscopic model is able to reproduce an important part of the available experimental data for $^{11}$Be and $^{19}$C.

The predictive power of the model  makes it particularly useful for exotic nuclei for which scarce information is available, such as $^{19}$C and other unknown nuclei. Although we compare with particle-rotor models, the proposed model does not need the \textit{core} to be neither a rotor nor a vibrator. That makes the model far more general and suitable for different regions of exotic nuclei. Whenever we have a certain knowledge of the \textit{core}, the model can be used to predict the spectrum of the \textit{core}+neutron (or \textit{core}+proton) composite system. Furthermore, since it uses the same THO formalism as the PRM model used in \cite{Mor12a}, it will be easily included in reaction calculations including \textit{core} excitations as we hope to show in future works.

 An important issue for extending the application of this model would be the correct application of the Pauli principle. Here we have removed those final deeply bound eigenstates that we considered as occupied by comparing with the spherical and Nilsson limits for each case. Differences between removing these eigenstates, removing pure spherical or Nilsson configurations, and more sophisticated treatments of the Pauli principle should lead to future developments of the present model.

In this work, particular nucleon-nucleon interaction (JLM) and method to extract core transition densities (AMD) have been used. However, the formalism proposed is general and not linked to them. It can be equally applied with any other appropriate NN interaction and/or model able to calculate core transition densities.

\begin{acknowledgments}
 This work has been partially supported by the Spanish Ministerio de 
 Ciencia e Innovaci\'on and FEDER funds under projects
 FIS2011-28738-c02-01,  FPA2009-07653, 
 FPA2009-08848 and  by the Spanish Consolider-Ingenio 2010 Programme CPAN
(CSD2007-00042)  and by Junta de  Andaluc\'ia (FQM160,
 P11-FQM-7632). J.A.L.\ acknowledges a research grant by the
 Ministerio de Ciencia e Innovaci\'on. 
\end{acknowledgments}

\bibliography{pamd}

\begin{thebibliography}{57}%
\makeatletter
\providecommand \@ifxundefined [1]{%
 \@ifx{#1\undefined}
}%
\providecommand \@ifnum [1]{%
 \ifnum #1\expandafter \@firstoftwo
 \else \expandafter \@secondoftwo
 \fi
}%
\providecommand \@ifx [1]{%
 \ifx #1\expandafter \@firstoftwo
 \else \expandafter \@secondoftwo
 \fi
}%
\providecommand \natexlab [1]{#1}%
\providecommand \enquote  [1]{``#1''}%
\providecommand \bibnamefont  [1]{#1}%
\providecommand \bibfnamefont [1]{#1}%
\providecommand \citenamefont [1]{#1}%
\providecommand \href@noop [0]{\@secondoftwo}%
\providecommand \href [0]{\begingroup \@sanitize@url \@href}%
\providecommand \@href[1]{\@@startlink{#1}\@@href}%
\providecommand \@@href[1]{\endgroup#1\@@endlink}%
\providecommand \@sanitize@url [0]{\catcode `\\12\catcode `\$12\catcode
  `\&12\catcode `\#12\catcode `\^12\catcode `\_12\catcode `\%12\relax}%
\providecommand \@@startlink[1]{}%
\providecommand \@@endlink[0]{}%
\providecommand \url  [0]{\begingroup\@sanitize@url \@url }%
\providecommand \@url [1]{\endgroup\@href {#1}{\urlprefix }}%
\providecommand \urlprefix  [0]{URL }%
\providecommand \Eprint [0]{\href }%
\providecommand \doibase [0]{http://dx.doi.org/}%
\providecommand \selectlanguage [0]{\@gobble}%
\providecommand \bibinfo  [0]{\@secondoftwo}%
\providecommand \bibfield  [0]{\@secondoftwo}%
\providecommand \translation [1]{[#1]}%
\providecommand \BibitemOpen [0]{}%
\providecommand \bibitemStop [0]{}%
\providecommand \bibitemNoStop [0]{.\EOS\space}%
\providecommand \EOS [0]{\spacefactor3000\relax}%
\providecommand \BibitemShut  [1]{\csname bibitem#1\endcsname}%
\let\auto@bib@innerbib\@empty
\bibitem [{\citenamefont {Austern}\ \emph {et~al.}(1987)\citenamefont
  {Austern}, \citenamefont {Iseri}, \citenamefont {Kamimura}, \citenamefont
  {Kawai}, \citenamefont {Rawitscher},\ and\ \citenamefont {Yahiro}}]{CDCC}%
  \BibitemOpen
  \bibfield  {author} {\bibinfo {author} {\bibfnamefont {N.}~\bibnamefont
  {Austern}}, \bibinfo {author} {\bibfnamefont {Y.}~\bibnamefont {Iseri}},
  \bibinfo {author} {\bibfnamefont {M.}~\bibnamefont {Kamimura}}, \bibinfo
  {author} {\bibfnamefont {M.}~\bibnamefont {Kawai}}, \bibinfo {author}
  {\bibfnamefont {G.}~\bibnamefont {Rawitscher}}, \ and\ \bibinfo {author}
  {\bibfnamefont {M.}~\bibnamefont {Yahiro}},\ }\href {\doibase
  10.1016/0370-1573(87)90094-9} {\bibfield  {journal} {\bibinfo  {journal}
  {Phys. Rep.}\ }\textbf {\bibinfo {volume} {154}},\ \bibinfo {pages} {125}
  (\bibinfo {year} {1987})}\BibitemShut {NoStop}%
\bibitem [{\citenamefont {Banerjee}\ and\ \citenamefont {Shyam}(2000)}]{Ban00}%
  \BibitemOpen
  \bibfield  {author} {\bibinfo {author} {\bibfnamefont {P.}~\bibnamefont
  {Banerjee}}\ and\ \bibinfo {author} {\bibfnamefont {R.}~\bibnamefont
  {Shyam}},\ }\href {\doibase 10.1103/PhysRevC.61.047301} {\bibfield  {journal}
  {\bibinfo  {journal} {Phys. Rev. C}\ }\textbf {\bibinfo {volume} {61}},\
  \bibinfo {pages} {047301} (\bibinfo {year} {2000})}\BibitemShut {NoStop}%
\bibitem [{\citenamefont {Tostevin}\ \emph {et~al.}(1998)\citenamefont
  {Tostevin}, \citenamefont {Rugmai},\ and\ \citenamefont {Johnson}}]{Tos98}%
  \BibitemOpen
  \bibfield  {author} {\bibinfo {author} {\bibfnamefont {J.~A.}\ \bibnamefont
  {Tostevin}}, \bibinfo {author} {\bibfnamefont {S.}~\bibnamefont {Rugmai}}, \
  and\ \bibinfo {author} {\bibfnamefont {R.~C.}\ \bibnamefont {Johnson}},\
  }\href {\doibase 10.1103/PhysRevC.57.3225} {\bibfield  {journal} {\bibinfo
  {journal} {Phys. Rev. C}\ }\textbf {\bibinfo {volume} {57}},\ \bibinfo
  {pages} {3225} (\bibinfo {year} {1998})}\BibitemShut {NoStop}%
\bibitem [{\citenamefont {Faddeev}(1960)}]{faddeev60}%
  \BibitemOpen
  \bibfield  {author} {\bibinfo {author} {\bibfnamefont {L.~D.}\ \bibnamefont
  {Faddeev}},\ }\href@noop {} {\bibfield  {journal} {\bibinfo  {journal} {Zh.
  Eksp. Theor. Fiz.}\ }\textbf {\bibinfo {volume} {39}},\ \bibinfo {pages}
  {1459} (\bibinfo {year} {1960})},\ \bibinfo {note} {{[Sov. Phys. JETP {\bf
  12}, 1014 (1961)]}}\BibitemShut {NoStop}%
\bibitem [{\citenamefont {Alt}\ \emph {et~al.}(1967)\citenamefont {Alt},
  \citenamefont {Grassberger},\ and\ \citenamefont {Sandhas}}]{Alt}%
  \BibitemOpen
  \bibfield  {author} {\bibinfo {author} {\bibfnamefont {E.~O.}\ \bibnamefont
  {Alt}}, \bibinfo {author} {\bibfnamefont {P.}~\bibnamefont {Grassberger}}, \
  and\ \bibinfo {author} {\bibfnamefont {W.}~\bibnamefont {Sandhas}},\ }\href
  {\doibase 10.1016/0550-3213(67)90016-8} {\bibfield  {journal} {\bibinfo
  {journal} {Nucl. Phys.}\ }\textbf {\bibinfo {volume} {B2}},\ \bibinfo {pages}
  {167} (\bibinfo {year} {1967})}\BibitemShut {NoStop}%
\bibitem [{\citenamefont {{Typel}}\ and\ \citenamefont {{Baur}}(1994)}]{Typ94}%
  \BibitemOpen
  \bibfield  {author} {\bibinfo {author} {\bibfnamefont {S.}~\bibnamefont
  {{Typel}}}\ and\ \bibinfo {author} {\bibfnamefont {G.}~\bibnamefont
  {{Baur}}},\ }\href {\doibase 10.1103/PhysRevC.50.2104} {\bibfield  {journal}
  {\bibinfo  {journal} {Phys.\ Rev. C}\ }\textbf {\bibinfo {volume} {50}},\
  \bibinfo {pages} {2104} (\bibinfo {year} {1994})}\BibitemShut {NoStop}%
\bibitem [{\citenamefont {Esbensen}\ and\ \citenamefont
  {Bertsch}(1996)}]{Esb96}%
  \BibitemOpen
  \bibfield  {author} {\bibinfo {author} {\bibfnamefont {H.}~\bibnamefont
  {Esbensen}}\ and\ \bibinfo {author} {\bibfnamefont {G.~F.}\ \bibnamefont
  {Bertsch}},\ }\href {\doibase 10.1016/0375-9474(96)00006-1} {\bibfield
  {journal} {\bibinfo  {journal} {Nucl. Phys.}\ }\textbf {\bibinfo {volume}
  {A600}},\ \bibinfo {pages} {37} (\bibinfo {year} {1996})}\BibitemShut
  {NoStop}%
\bibitem [{\citenamefont {Kido}\ \emph {et~al.}(1994)\citenamefont {Kido},
  \citenamefont {Yabana},\ and\ \citenamefont {Suzuki}}]{Kid94}%
  \BibitemOpen
  \bibfield  {author} {\bibinfo {author} {\bibfnamefont {T.}~\bibnamefont
  {Kido}}, \bibinfo {author} {\bibfnamefont {K.}~\bibnamefont {Yabana}}, \ and\
  \bibinfo {author} {\bibfnamefont {Y.}~\bibnamefont {Suzuki}},\ }\href
  {\doibase 10.1103/PhysRevC.50.R1276} {\bibfield  {journal} {\bibinfo
  {journal} {Phys. Rev. C}\ }\textbf {\bibinfo {volume} {50}},\ \bibinfo
  {pages} {R1276} (\bibinfo {year} {1994})}\BibitemShut {NoStop}%
\bibitem [{\citenamefont {{Typel}}\ and\ \citenamefont {{Baur}}(2001)}]{Typ01}%
  \BibitemOpen
  \bibfield  {author} {\bibinfo {author} {\bibfnamefont {S.}~\bibnamefont
  {{Typel}}}\ and\ \bibinfo {author} {\bibfnamefont {G.}~\bibnamefont
  {{Baur}}},\ }\href {\doibase 10.1103/PhysRevC.64.024601} {\bibfield
  {journal} {\bibinfo  {journal} {Phys.\ Rev. C}\ }\textbf {\bibinfo {volume}
  {64}},\ \bibinfo {pages} {024601} (\bibinfo {year} {2001})}\BibitemShut
  {NoStop}%
\bibitem [{\citenamefont {Capel}\ \emph {et~al.}(2004)\citenamefont {Capel},
  \citenamefont {Goldstein},\ and\ \citenamefont {Baye}}]{Cap04}%
  \BibitemOpen
  \bibfield  {author} {\bibinfo {author} {\bibfnamefont {P.}~\bibnamefont
  {Capel}}, \bibinfo {author} {\bibfnamefont {G.}~\bibnamefont {Goldstein}}, \
  and\ \bibinfo {author} {\bibfnamefont {D.}~\bibnamefont {Baye}},\ }\href
  {\doibase 10.1103/PhysRevC.70.064605} {\bibfield  {journal} {\bibinfo
  {journal} {Phys. Rev. C}\ }\textbf {\bibinfo {volume} {70}},\ \bibinfo
  {pages} {064605} (\bibinfo {year} {2004})}\BibitemShut {NoStop}%
\bibitem [{\citenamefont {{Garcia-Camacho}}\ \emph {et~al.}(2006)\citenamefont
  {{Garcia-Camacho}}, \citenamefont {{Bonaccorso}},\ and\ \citenamefont
  {{Brink}}}]{Gar06}%
  \BibitemOpen
  \bibfield  {author} {\bibinfo {author} {\bibfnamefont {A.}~\bibnamefont
  {{Garcia-Camacho}}}, \bibinfo {author} {\bibfnamefont {A.}~\bibnamefont
  {{Bonaccorso}}}, \ and\ \bibinfo {author} {\bibfnamefont {D.~M.}\
  \bibnamefont {{Brink}}},\ }\href {\doibase 10.1016/j.nuclphysa.2006.07.033}
  {\bibfield  {journal} {\bibinfo  {journal} {Nucl. Phys.}\ }\textbf {\bibinfo
  {volume} {A776}},\ \bibinfo {pages} {118} (\bibinfo {year}
  {2006})}\BibitemShut {NoStop}%
\bibitem [{\citenamefont {Bohr}\ and\ \citenamefont {Mottelson}(1969)}]{BM}%
  \BibitemOpen
  \bibfield  {author} {\bibinfo {author} {\bibfnamefont {A.}~\bibnamefont
  {Bohr}}\ and\ \bibinfo {author} {\bibfnamefont {B.}~\bibnamefont
  {Mottelson}},\ }\href@noop {} {\emph {\bibinfo {title} {Nuclear
  Structure}}},\ \bibinfo {edition} {{New York, W. A. Benjamin}}\ ed.\
  (\bibinfo {year} {1969})\BibitemShut {NoStop}%
\bibitem [{\citenamefont {{Vinh Mau}}(1995)}]{Vin95}%
  \BibitemOpen
  \bibfield  {author} {\bibinfo {author} {\bibfnamefont {N.}~\bibnamefont
  {{Vinh Mau}}},\ }\href {\doibase 10.1016/0375-9474(95)00298-F} {\bibfield
  {journal} {\bibinfo  {journal} {Nucl. Phys.}\ }\textbf {\bibinfo {volume}
  {A592}},\ \bibinfo {pages} {33} (\bibinfo {year} {1995})}\BibitemShut
  {NoStop}%
\bibitem [{\citenamefont {{Gori}}\ \emph {et~al.}(2004)\citenamefont {{Gori}},
  \citenamefont {{Barranco}}, \citenamefont {{Vigezzi}},\ and\ \citenamefont
  {{Broglia}}}]{Gor04}%
  \BibitemOpen
  \bibfield  {author} {\bibinfo {author} {\bibfnamefont {G.}~\bibnamefont
  {{Gori}}}, \bibinfo {author} {\bibfnamefont {F.}~\bibnamefont {{Barranco}}},
  \bibinfo {author} {\bibfnamefont {E.}~\bibnamefont {{Vigezzi}}}, \ and\
  \bibinfo {author} {\bibfnamefont {R.~A.}\ \bibnamefont {{Broglia}}},\ }\href
  {\doibase 10.1103/PhysRevC.69.041302} {\bibfield  {journal} {\bibinfo
  {journal} {Phys.Rev. C}\ }\textbf {\bibinfo {volume} {69}},\ \bibinfo {pages}
  {041302} (\bibinfo {year} {2004})}\BibitemShut {NoStop}%
\bibitem [{\citenamefont {Descouvemont}\ and\ \citenamefont
  {Hussein}(2013)}]{Des13}%
  \BibitemOpen
  \bibfield  {author} {\bibinfo {author} {\bibfnamefont {P.}~\bibnamefont
  {Descouvemont}}\ and\ \bibinfo {author} {\bibfnamefont {M.~S.}\ \bibnamefont
  {Hussein}},\ }\href {\doibase 10.1103/PhysRevLett.111.082701} {\bibfield
  {journal} {\bibinfo  {journal} {Phys. Rev. Lett.}\ }\textbf {\bibinfo
  {volume} {111}},\ \bibinfo {pages} {082701} (\bibinfo {year}
  {2013})}\BibitemShut {NoStop}%
\bibitem [{\citenamefont {Satchler}\ and\ \citenamefont {Love}(1979)}]{Sat79}%
  \BibitemOpen
  \bibfield  {author} {\bibinfo {author} {\bibfnamefont {G.~R.}\ \bibnamefont
  {Satchler}}\ and\ \bibinfo {author} {\bibfnamefont {W.~G.}\ \bibnamefont
  {Love}},\ }\href {\doibase 10.1016/0370-1573(79)90081-4} {\bibfield
  {journal} {\bibinfo  {journal} {Phys. Rep.}\ }\textbf {\bibinfo {volume}
  {55}},\ \bibinfo {pages} {183} (\bibinfo {year} {1979})}\BibitemShut
  {NoStop}%
\bibitem [{\citenamefont {Jeukenne}\ \emph {et~al.}(1977)\citenamefont
  {Jeukenne}, \citenamefont {Lejeune},\ and\ \citenamefont {Mahaux}}]{JLM}%
  \BibitemOpen
  \bibfield  {author} {\bibinfo {author} {\bibfnamefont {J.-P.}\ \bibnamefont
  {Jeukenne}}, \bibinfo {author} {\bibfnamefont {A.}~\bibnamefont {Lejeune}}, \
  and\ \bibinfo {author} {\bibfnamefont {C.}~\bibnamefont {Mahaux}},\ }\href
  {\doibase 10.1103/PhysRevC.16.80} {\bibfield  {journal} {\bibinfo  {journal}
  {Phys. Rev. C}\ }\textbf {\bibinfo {volume} {16}},\ \bibinfo {pages} {80}
  (\bibinfo {year} {1977})}\BibitemShut {NoStop}%
\bibitem [{\citenamefont {{Kanada-Enyo}}\ \emph {et~al.}(1995)\citenamefont
  {{Kanada-Enyo}}, \citenamefont {{Horiuchi}},\ and\ \citenamefont
  {{Ono}}}]{Kan95a}%
  \BibitemOpen
  \bibfield  {author} {\bibinfo {author} {\bibfnamefont {Y.}~\bibnamefont
  {{Kanada-Enyo}}}, \bibinfo {author} {\bibfnamefont {H.}~\bibnamefont
  {{Horiuchi}}}, \ and\ \bibinfo {author} {\bibfnamefont {A.}~\bibnamefont
  {{Ono}}},\ }\href {\doibase 10.1103/PhysRevC.52.628} {\bibfield  {journal}
  {\bibinfo  {journal} {Phys. Rev. C}\ }\textbf {\bibinfo {volume} {52}},\
  \bibinfo {pages} {628} (\bibinfo {year} {1995})}\BibitemShut {NoStop}%
\bibitem [{\citenamefont {{Kanada-Enyo}}\ and\ \citenamefont
  {{Horiuchi}}(1995)}]{Kan95b}%
  \BibitemOpen
  \bibfield  {author} {\bibinfo {author} {\bibfnamefont {Y.}~\bibnamefont
  {{Kanada-Enyo}}}\ and\ \bibinfo {author} {\bibfnamefont {H.}~\bibnamefont
  {{Horiuchi}}},\ }\href {\doibase 10.1103/PhysRevC.52.647} {\bibfield
  {journal} {\bibinfo  {journal} {Phys. Rev. C}\ }\textbf {\bibinfo {volume}
  {52}},\ \bibinfo {pages} {647} (\bibinfo {year} {1995})}\BibitemShut
  {NoStop}%
\bibitem [{\citenamefont {{Karataglidis}}\ \emph {et~al.}(2005)\citenamefont
  {{Karataglidis}}, \citenamefont {{Amos}},\ and\ \citenamefont
  {{Giraud}}}]{Amos}%
  \BibitemOpen
  \bibfield  {author} {\bibinfo {author} {\bibfnamefont {S.}~\bibnamefont
  {{Karataglidis}}}, \bibinfo {author} {\bibfnamefont {K.}~\bibnamefont
  {{Amos}}}, \ and\ \bibinfo {author} {\bibfnamefont {B.~G.}\ \bibnamefont
  {{Giraud}}},\ }\href {\doibase 10.1103/PhysRevC.71.064601} {\bibfield
  {journal} {\bibinfo  {journal} {Phys. Rev. C}\ }\textbf {\bibinfo {volume}
  {71}},\ \bibinfo {pages} {064601} (\bibinfo {year} {2005})}\BibitemShut
  {NoStop}%
\bibitem [{\citenamefont {{Lay}}\ \emph {et~al.}(2010)\citenamefont {{Lay}},
  \citenamefont {{Moro}}, \citenamefont {{Arias}},\ and\ \citenamefont
  {{Gomez-Camacho}}}]{Lay10}%
  \BibitemOpen
  \bibfield  {author} {\bibinfo {author} {\bibfnamefont {J.~A.}\ \bibnamefont
  {{Lay}}}, \bibinfo {author} {\bibfnamefont {A.~M.}\ \bibnamefont {{Moro}}},
  \bibinfo {author} {\bibfnamefont {J.~M.}\ \bibnamefont {{Arias}}}, \ and\
  \bibinfo {author} {\bibfnamefont {J.}~\bibnamefont {{Gomez-Camacho}}},\
  }\href {\doibase 10.1103/PhysRevC.82.024605} {\bibfield  {journal} {\bibinfo
  {journal} {Phys. Rev. C}\ }\textbf {\bibinfo {volume} {82}},\ \bibinfo
  {pages} {024605} (\bibinfo {year} {2010})}\BibitemShut {NoStop}%
\bibitem [{\citenamefont {Lay}\ \emph {et~al.}(2012{\natexlab{a}})\citenamefont
  {Lay}, \citenamefont {Moro}, \citenamefont {Arias},\ and\ \citenamefont
  {G\'omez-Camacho}}]{Lay12}%
  \BibitemOpen
  \bibfield  {author} {\bibinfo {author} {\bibfnamefont {J.~A.}\ \bibnamefont
  {Lay}}, \bibinfo {author} {\bibfnamefont {A.~M.}\ \bibnamefont {Moro}},
  \bibinfo {author} {\bibfnamefont {J.~M.}\ \bibnamefont {Arias}}, \ and\
  \bibinfo {author} {\bibfnamefont {J.}~\bibnamefont {G\'omez-Camacho}},\
  }\href {\doibase 10.1103/PhysRevC.85.054618} {\bibfield  {journal} {\bibinfo
  {journal} {Phys. Rev. C}\ }\textbf {\bibinfo {volume} {85}},\ \bibinfo
  {pages} {054618} (\bibinfo {year} {2012}{\natexlab{a}})}\BibitemShut
  {NoStop}%
\bibitem [{\citenamefont {Hazi}\ and\ \citenamefont {Taylor}(1970)}]{Haz70}%
  \BibitemOpen
  \bibfield  {author} {\bibinfo {author} {\bibfnamefont {A.~U.}\ \bibnamefont
  {Hazi}}\ and\ \bibinfo {author} {\bibfnamefont {H.~S.}\ \bibnamefont
  {Taylor}},\ }\href {\doibase 10.1103/PhysRevA.1.1109} {\bibfield  {journal}
  {\bibinfo  {journal} {Phys. Rev. A}\ }\textbf {\bibinfo {volume} {1}},\
  \bibinfo {pages} {1109} (\bibinfo {year} {1970})}\BibitemShut {NoStop}%
\bibitem [{\citenamefont {Lay}\ \emph {et~al.}(2012{\natexlab{b}})\citenamefont
  {Lay}, \citenamefont {Arias}, \citenamefont {G\'omez-Camacho},\ and\
  \citenamefont {Moro}}]{Lay12proc}%
  \BibitemOpen
  \bibfield  {author} {\bibinfo {author} {\bibfnamefont {J.~A.}\ \bibnamefont
  {Lay}}, \bibinfo {author} {\bibfnamefont {J.~M.}\ \bibnamefont {Arias}},
  \bibinfo {author} {\bibfnamefont {J.}~\bibnamefont {G\'omez-Camacho}}, \ and\
  \bibinfo {author} {\bibfnamefont {A.~M.}\ \bibnamefont {Moro}},\ }\href
  {\doibase 10.1063/1.4759428} {\bibfield  {journal} {\bibinfo  {journal} {AIP
  Conf. Proc.}\ }\textbf {\bibinfo {volume} {1488}},\ \bibinfo {pages} {436}
  (\bibinfo {year} {2012}{\natexlab{b}})}\BibitemShut {NoStop}%
\bibitem [{\citenamefont {Brink}\ and\ \citenamefont {{Satchler}}(1968)}]{BS}%
  \BibitemOpen
  \bibfield  {author} {\bibinfo {author} {\bibfnamefont {D.~M.}\ \bibnamefont
  {Brink}}\ and\ \bibinfo {author} {\bibfnamefont {G.~R.}\ \bibnamefont
  {{Satchler}}},\ }\href@noop {} {\emph {\bibinfo {title} {Angular Momentum}}}\
  (\bibinfo  {publisher} {Clarendon, Oxford},\ \bibinfo {year}
  {1968})\BibitemShut {NoStop}%
\bibitem [{\citenamefont {Tamura}(1965)}]{Tam65}%
  \BibitemOpen
  \bibfield  {author} {\bibinfo {author} {\bibfnamefont {T.}~\bibnamefont
  {Tamura}},\ }\href {\doibase 10.1103/RevModPhys.37.679} {\bibfield  {journal}
  {\bibinfo  {journal} {Rev. Mod. Phys.}\ }\textbf {\bibinfo {volume} {37}},\
  \bibinfo {pages} {679} (\bibinfo {year} {1965})}\BibitemShut {NoStop}%
\bibitem [{\citenamefont {Thompson}\ and\ \citenamefont {Nunes}(2009)}]{Tho09}%
  \BibitemOpen
  \bibfield  {author} {\bibinfo {author} {\bibfnamefont {I.~J.}\ \bibnamefont
  {Thompson}}\ and\ \bibinfo {author} {\bibfnamefont {F.~M.}\ \bibnamefont
  {Nunes}},\ }\href@noop {} {\emph {\bibinfo {title} {Nuclear reactions for
  astrophysics}}}\ (\bibinfo  {publisher} {Cambridge, UK: Cambridge University
  Press},\ \bibinfo {year} {2009})\BibitemShut {NoStop}%
\bibitem [{\citenamefont {Kanada-En'yo}\ \emph {et~al.}(1999)\citenamefont
  {Kanada-En'yo}, \citenamefont {Horiuchi},\ and\ \citenamefont
  {Dot{\'e}}}]{Kan99}%
  \BibitemOpen
  \bibfield  {author} {\bibinfo {author} {\bibfnamefont {Y.}~\bibnamefont
  {Kanada-En'yo}}, \bibinfo {author} {\bibfnamefont {H.}~\bibnamefont
  {Horiuchi}}, \ and\ \bibinfo {author} {\bibfnamefont {A.}~\bibnamefont
  {Dot{\'e}}},\ }\href {\doibase 10.1103/PhysRevC.60.064304} {\bibfield
  {journal} {\bibinfo  {journal} {Phys. Rev. C}\ }\textbf {\bibinfo {volume}
  {60}},\ \bibinfo {pages} {064304} (\bibinfo {year} {1999})}\BibitemShut
  {NoStop}%
\bibitem [{\citenamefont {Kanada-En'yo}\ \emph
  {et~al.}(2013{\natexlab{a}})\citenamefont {Kanada-En'yo}, \citenamefont
  {Kobayashi},\ and\ \citenamefont {Suhara}}]{Kan13a}%
  \BibitemOpen
  \bibfield  {author} {\bibinfo {author} {\bibfnamefont {Y.}~\bibnamefont
  {Kanada-En'yo}}, \bibinfo {author} {\bibfnamefont {F.}~\bibnamefont
  {Kobayashi}}, \ and\ \bibinfo {author} {\bibfnamefont {T.}~\bibnamefont
  {Suhara}},\ }\href {\doibase 10.1088/1742-6596/436/1/012037} {\bibfield
  {journal} {\bibinfo  {journal} {Journal of Physics: Conference Series}\
  }\textbf {\bibinfo {volume} {436}},\ \bibinfo {pages} {012037} (\bibinfo
  {year} {2013}{\natexlab{a}})}\BibitemShut {NoStop}%
\bibitem [{\citenamefont {Kanada-En'yo}\ \emph
  {et~al.}(2013{\natexlab{b}})\citenamefont {Kanada-En'yo}, \citenamefont
  {Kobayashi},\ and\ \citenamefont {Suhara}}]{Kan13b}%
  \BibitemOpen
  \bibfield  {author} {\bibinfo {author} {\bibfnamefont {Y.}~\bibnamefont
  {Kanada-En'yo}}, \bibinfo {author} {\bibfnamefont {F.}~\bibnamefont
  {Kobayashi}}, \ and\ \bibinfo {author} {\bibfnamefont {T.}~\bibnamefont
  {Suhara}},\ }\href {\doibase 10.1088/1742-6596/445/1/012037} {\bibfield
  {journal} {\bibinfo  {journal} {Journal of Physics: Conference Series}\
  }\textbf {\bibinfo {volume} {445}},\ \bibinfo {pages} {012037} (\bibinfo
  {year} {2013}{\natexlab{b}})}\BibitemShut {NoStop}%
\bibitem [{\citenamefont {von Oertzen}(1996)}]{Oer96}%
  \BibitemOpen
  \bibfield  {author} {\bibinfo {author} {\bibfnamefont {W.}~\bibnamefont {von
  Oertzen}},\ }\href {\doibase 10.1007/s002180050010} {\bibfield  {journal}
  {\bibinfo  {journal} {Z. Phys. A}\ }\textbf {\bibinfo {volume} {354}},\
  \bibinfo {pages} {37} (\bibinfo {year} {1996})}\BibitemShut {NoStop}%
\bibitem [{\citenamefont {Takashina}\ \emph {et~al.}(2005)\citenamefont
  {Takashina}, \citenamefont {Kanada-En'yo},\ and\ \citenamefont
  {Sakuragi}}]{Tak05}%
  \BibitemOpen
  \bibfield  {author} {\bibinfo {author} {\bibfnamefont {M.}~\bibnamefont
  {Takashina}}, \bibinfo {author} {\bibfnamefont {Y.}~\bibnamefont
  {Kanada-En'yo}}, \ and\ \bibinfo {author} {\bibfnamefont {Y.}~\bibnamefont
  {Sakuragi}},\ }\href {\doibase 10.1103/PhysRevC.71.054602} {\bibfield
  {journal} {\bibinfo  {journal} {Phys. Rev. C}\ }\textbf {\bibinfo {volume}
  {71}},\ \bibinfo {pages} {054602} (\bibinfo {year} {2005})}\BibitemShut
  {NoStop}%
\bibitem [{\citenamefont {Takashina}\ and\ \citenamefont
  {Kanada-En'yo}(2008)}]{Tak08}%
  \BibitemOpen
  \bibfield  {author} {\bibinfo {author} {\bibfnamefont {M.}~\bibnamefont
  {Takashina}}\ and\ \bibinfo {author} {\bibfnamefont {Y.}~\bibnamefont
  {Kanada-En'yo}},\ }\href {\doibase 10.1103/PhysRevC.77.014604} {\bibfield
  {journal} {\bibinfo  {journal} {Phys. Rev. C}\ }\textbf {\bibinfo {volume}
  {77}},\ \bibinfo {pages} {014604} (\bibinfo {year} {2008})}\BibitemShut
  {NoStop}%
\bibitem [{\citenamefont {Tarutina}\ and\ \citenamefont
  {Hussein}(2004)}]{Tar04}%
  \BibitemOpen
  \bibfield  {author} {\bibinfo {author} {\bibfnamefont {T.}~\bibnamefont
  {Tarutina}}\ and\ \bibinfo {author} {\bibfnamefont {M.~S.}\ \bibnamefont
  {Hussein}},\ }\href {\doibase 10.1103/PhysRevC.70.034603} {\bibfield
  {journal} {\bibinfo  {journal} {Phys. Rev. C}\ }\textbf {\bibinfo {volume}
  {70}},\ \bibinfo {pages} {034603} (\bibinfo {year} {2004})}\BibitemShut
  {NoStop}%
\bibitem [{\citenamefont {Nunes}\ \emph
  {et~al.}(1996{\natexlab{a}})\citenamefont {Nunes}, \citenamefont {Christley},
  \citenamefont {Thompson}, \citenamefont {Johnson},\ and\ \citenamefont
  {Efros}}]{Nun96}%
  \BibitemOpen
  \bibfield  {author} {\bibinfo {author} {\bibfnamefont {F.~M.}\ \bibnamefont
  {Nunes}}, \bibinfo {author} {\bibfnamefont {J.~A.}\ \bibnamefont
  {Christley}}, \bibinfo {author} {\bibfnamefont {I.~J.}\ \bibnamefont
  {Thompson}}, \bibinfo {author} {\bibfnamefont {R.~C.}\ \bibnamefont
  {Johnson}}, \ and\ \bibinfo {author} {\bibfnamefont {V.~D.}\ \bibnamefont
  {Efros}},\ }\href {\doibase 10.1016/0375-9474(96)00284-9} {\bibfield
  {journal} {\bibinfo  {journal} {Nucl. Phys.}\ }\textbf {\bibinfo {volume}
  {A609}},\ \bibinfo {pages} {43} (\bibinfo {year}
  {1996}{\natexlab{a}})}\BibitemShut {NoStop}%
\bibitem [{\citenamefont {{Tarutina}}\ \emph {et~al.}(2006)\citenamefont
  {{Tarutina}}, \citenamefont {{Samana}}, \citenamefont {{Krmpotic}},\ and\
  \citenamefont {{Hussein}}}]{Tar06}%
  \BibitemOpen
  \bibfield  {author} {\bibinfo {author} {\bibfnamefont {T.}~\bibnamefont
  {{Tarutina}}}, \bibinfo {author} {\bibfnamefont {A.~R.}\ \bibnamefont
  {{Samana}}}, \bibinfo {author} {\bibfnamefont {F.}~\bibnamefont
  {{Krmpotic}}}, \ and\ \bibinfo {author} {\bibfnamefont {M.~S.}\ \bibnamefont
  {{Hussein}}},\ }\href@noop {} {\bibfield  {journal} {\bibinfo  {journal}
  {Braz. J. Phys.}\ }\textbf {\bibinfo {volume} {36}},\ \bibinfo {pages} {1349}
  (\bibinfo {year} {2006})}\BibitemShut {NoStop}%
\bibitem [{\citenamefont {Nunes}\ \emph
  {et~al.}(1996{\natexlab{b}})\citenamefont {Nunes}, \citenamefont {Thompson},\
  and\ \citenamefont {Johnson}}]{Nun96a}%
  \BibitemOpen
  \bibfield  {author} {\bibinfo {author} {\bibfnamefont {F.~M.}\ \bibnamefont
  {Nunes}}, \bibinfo {author} {\bibfnamefont {I.~J.}\ \bibnamefont {Thompson}},
  \ and\ \bibinfo {author} {\bibfnamefont {R.~C.}\ \bibnamefont {Johnson}},\
  }\href {\doibase http://dx.doi.org/10.1016/0375-9474(95)00398-3} {\bibfield
  {journal} {\bibinfo  {journal} {Nucl. Phys.}\ }\textbf {\bibinfo {volume}
  {A596}},\ \bibinfo {pages} {171 } (\bibinfo {year}
  {1996}{\natexlab{b}})}\BibitemShut {NoStop}%
\bibitem [{\citenamefont {Takashina}()}]{MINC}%
  \BibitemOpen
  \bibfield  {author} {\bibinfo {author} {\bibfnamefont {M.}~\bibnamefont
  {Takashina}},\ }\href@noop {} {}\bibinfo {note} {Single-folding model code
  {\sc MINC}. Private Comunication}\BibitemShut {NoStop}%
\bibitem [{\citenamefont {Fukuda}\ \emph {et~al.}(2004)\citenamefont {Fukuda}
  \emph {et~al.}}]{Fuk04}%
  \BibitemOpen
  \bibfield  {author} {\bibinfo {author} {\bibfnamefont {N.}~\bibnamefont
  {Fukuda}} \emph {et~al.},\ }\href {\doibase 10.1103/PhysRevC.70.054606}
  {\bibfield  {journal} {\bibinfo  {journal} {Phys. Rev. C}\ }\textbf {\bibinfo
  {volume} {70}},\ \bibinfo {pages} {054606} (\bibinfo {year}
  {2004})}\BibitemShut {NoStop}%
\bibitem [{\citenamefont {Hirayama}\ \emph {et~al.}(2005)\citenamefont
  {Hirayama}, \citenamefont {Shimoda}, \citenamefont {Izumi}, \citenamefont
  {Hatakeyama}, \citenamefont {Jackson}, \citenamefont {Levy}, \citenamefont
  {Miyatake}, \citenamefont {Yagi},\ and\ \citenamefont {Yano}}]{Hir05}%
  \BibitemOpen
  \bibfield  {author} {\bibinfo {author} {\bibfnamefont {Y.}~\bibnamefont
  {Hirayama}}, \bibinfo {author} {\bibfnamefont {T.}~\bibnamefont {Shimoda}},
  \bibinfo {author} {\bibfnamefont {H.}~\bibnamefont {Izumi}}, \bibinfo
  {author} {\bibfnamefont {A.}~\bibnamefont {Hatakeyama}}, \bibinfo {author}
  {\bibfnamefont {K.~P.}\ \bibnamefont {Jackson}}, \bibinfo {author}
  {\bibfnamefont {C.~D.~P.}\ \bibnamefont {Levy}}, \bibinfo {author}
  {\bibfnamefont {H.}~\bibnamefont {Miyatake}}, \bibinfo {author}
  {\bibfnamefont {M.}~\bibnamefont {Yagi}}, \ and\ \bibinfo {author}
  {\bibfnamefont {H.}~\bibnamefont {Yano}},\ }\href {\doibase
  10.1016/j.physletb.2005.02.040} {\bibfield  {journal} {\bibinfo  {journal}
  {Phys. Lett. B}\ }\textbf {\bibinfo {volume} {611}},\ \bibinfo {pages} {239}
  (\bibinfo {year} {2005})}\BibitemShut {NoStop}%
\bibitem [{\citenamefont {Nunes}(1995)}]{Nun95}%
  \BibitemOpen
  \bibfield  {author} {\bibinfo {author} {\bibfnamefont {F.~M.}\ \bibnamefont
  {Nunes}},\ }\emph {\bibinfo {title} {{Core Excitation in Few Body Systems:
  Application to halo nuclei}}},\ \href@noop {} {Ph.D. thesis},\ \bibinfo
  {school} {University of Surrey} (\bibinfo {year} {1995})\BibitemShut
  {NoStop}%
\bibitem [{\citenamefont {Kukulin}\ \emph {et~al.}(1986)\citenamefont
  {Kukulin}, \citenamefont {Krasnopol'sky}, \citenamefont {Voronchev},\ and\
  \citenamefont {Sazonov}}]{Kuk86}%
  \BibitemOpen
  \bibfield  {author} {\bibinfo {author} {\bibfnamefont {V.~I.}\ \bibnamefont
  {Kukulin}}, \bibinfo {author} {\bibfnamefont {V.~M.}\ \bibnamefont
  {Krasnopol'sky}}, \bibinfo {author} {\bibfnamefont {V.~T.}\ \bibnamefont
  {Voronchev}}, \ and\ \bibinfo {author} {\bibfnamefont {P.~B.}\ \bibnamefont
  {Sazonov}},\ }\href {\doibase http://dx.doi.org/10.1016/0375-9474(86)90443-4}
  {\bibfield  {journal} {\bibinfo  {journal} {Nucl. Phys.}\ }\textbf {\bibinfo
  {volume} {A453}},\ \bibinfo {pages} {365 } (\bibinfo {year}
  {1986})}\BibitemShut {NoStop}%
\bibitem [{\citenamefont {Myo}\ \emph {et~al.}(2007)\citenamefont {Myo},
  \citenamefont {Kato}, \citenamefont {Toki},\ and\ \citenamefont
  {Ikeda}}]{Mio07}%
  \BibitemOpen
  \bibfield  {author} {\bibinfo {author} {\bibfnamefont {T.}~\bibnamefont
  {Myo}}, \bibinfo {author} {\bibfnamefont {K.}~\bibnamefont {Kato}}, \bibinfo
  {author} {\bibfnamefont {H.}~\bibnamefont {Toki}}, \ and\ \bibinfo {author}
  {\bibfnamefont {K.}~\bibnamefont {Ikeda}},\ }\href {\doibase
  10.1103/PhysRevC.76.024305} {\bibfield  {journal} {\bibinfo  {journal} {Phys.
  Rev. C}\ }\textbf {\bibinfo {volume} {76}},\ \bibinfo {pages} {024305}
  (\bibinfo {year} {2007})}\BibitemShut {NoStop}%
\bibitem [{\citenamefont {Warburton}\ and\ \citenamefont {Brown}(1992)}]{WBT}%
  \BibitemOpen
  \bibfield  {author} {\bibinfo {author} {\bibfnamefont {E.~K.}\ \bibnamefont
  {Warburton}}\ and\ \bibinfo {author} {\bibfnamefont {B.~A.}\ \bibnamefont
  {Brown}},\ }\href {\doibase 10.1103/PhysRevC.46.923} {\bibfield  {journal}
  {\bibinfo  {journal} {Phys. Rev. C}\ }\textbf {\bibinfo {volume} {46}},\
  \bibinfo {pages} {923} (\bibinfo {year} {1992})}\BibitemShut {NoStop}%
\bibitem [{\citenamefont {Brown}\ \emph {et~al.}(1986)\citenamefont {Brown},
  \citenamefont {Etchegoyen},\ and\ \citenamefont {Rae}}]{OXB}%
  \BibitemOpen
  \bibfield  {author} {\bibinfo {author} {\bibfnamefont {B.~A.}\ \bibnamefont
  {Brown}}, \bibinfo {author} {\bibfnamefont {A.}~\bibnamefont {Etchegoyen}}, \
  and\ \bibinfo {author} {\bibfnamefont {W.~D.~M.}\ \bibnamefont {Rae}},\
  }\href@noop {} {\emph {\bibinfo {title} {OXBASH-MSU (the
  Oxford--Buenos-Aires--Michigan State University shell model code)}}}\
  (\bibinfo  {publisher} {MSU-NSCL Report No. 524},\ \bibinfo {year}
  {1986})\BibitemShut {NoStop}%
\bibitem [{\citenamefont {Iwasaki}\ \emph {et~al.}(2000)\citenamefont {Iwasaki}
  \emph {et~al.}}]{Iwa00a}%
  \BibitemOpen
  \bibfield  {author} {\bibinfo {author} {\bibfnamefont {H.}~\bibnamefont
  {Iwasaki}} \emph {et~al.},\ }\href {\doibase 10.1016/S0370-2693(00)00428-7}
  {\bibfield  {journal} {\bibinfo  {journal} {Physics Letters B}\ }\textbf
  {\bibinfo {volume} {481}},\ \bibinfo {pages} {7 } (\bibinfo {year}
  {2000})}\BibitemShut {NoStop}%
\bibitem [{\citenamefont {Nakamura}\ \emph {et~al.}(1999)\citenamefont
  {Nakamura} \emph {et~al.}}]{Nak99}%
  \BibitemOpen
  \bibfield  {author} {\bibinfo {author} {\bibfnamefont {T.}~\bibnamefont
  {Nakamura}} \emph {et~al.},\ }\href {\doibase 10.1103/PhysRevLett.83.1112}
  {\bibfield  {journal} {\bibinfo  {journal} {Phys. Rev. Lett.}\ }\textbf
  {\bibinfo {volume} {83}},\ \bibinfo {pages} {1112} (\bibinfo {year}
  {1999})}\BibitemShut {NoStop}%
\bibitem [{\citenamefont {Typel}\ and\ \citenamefont {Shyam}(2001)}]{Typ01b}%
  \BibitemOpen
  \bibfield  {author} {\bibinfo {author} {\bibfnamefont {S.}~\bibnamefont
  {Typel}}\ and\ \bibinfo {author} {\bibfnamefont {R.}~\bibnamefont {Shyam}},\
  }\href {\doibase 10.1103/PhysRevC.64.024605} {\bibfield  {journal} {\bibinfo
  {journal} {Phys. Rev. C}\ }\textbf {\bibinfo {volume} {64}},\ \bibinfo
  {pages} {024605} (\bibinfo {year} {2001})}\BibitemShut {NoStop}%
\bibitem [{\citenamefont {Hamamoto}(2007)}]{Ham07}%
  \BibitemOpen
  \bibfield  {author} {\bibinfo {author} {\bibfnamefont {I.}~\bibnamefont
  {Hamamoto}},\ }\href {\doibase 10.1103/PhysRevC.76.054319} {\bibfield
  {journal} {\bibinfo  {journal} {Phys. Rev. C}\ }\textbf {\bibinfo {volume}
  {76}},\ \bibinfo {pages} {054319} (\bibinfo {year} {2007})}\BibitemShut
  {NoStop}%
\bibitem [{\citenamefont {Satou}\ \emph {et~al.}(2008)\citenamefont {Satou}
  \emph {et~al.}}]{Sat08}%
  \BibitemOpen
  \bibfield  {author} {\bibinfo {author} {\bibfnamefont {Y.}~\bibnamefont
  {Satou}} \emph {et~al.},\ }\href {\doibase 10.1016/j.physletb.2008.01.022}
  {\bibfield  {journal} {\bibinfo  {journal} {Phys. Lett. B}\ }\textbf
  {\bibinfo {volume} {660}},\ \bibinfo {pages} {320} (\bibinfo {year}
  {2008})}\BibitemShut {NoStop}%
\bibitem [{\citenamefont {Karataglidis}\ \emph {et~al.}(2008)\citenamefont
  {Karataglidis}, \citenamefont {Amos}, \citenamefont {Fraser}, \citenamefont
  {Canton},\ and\ \citenamefont {Svenne}}]{Kar08}%
  \BibitemOpen
  \bibfield  {author} {\bibinfo {author} {\bibfnamefont {S.}~\bibnamefont
  {Karataglidis}}, \bibinfo {author} {\bibfnamefont {K.}~\bibnamefont {Amos}},
  \bibinfo {author} {\bibfnamefont {P.}~\bibnamefont {Fraser}}, \bibinfo
  {author} {\bibfnamefont {L.}~\bibnamefont {Canton}}, \ and\ \bibinfo {author}
  {\bibfnamefont {J.}~\bibnamefont {Svenne}},\ }\href {\doibase
  10.1016/j.nuclphysa.2008.09.007} {\bibfield  {journal} {\bibinfo  {journal}
  {Nucl. Phys.}\ }\textbf {\bibinfo {volume} {A813}},\ \bibinfo {pages} {235}
  (\bibinfo {year} {2008})}\BibitemShut {NoStop}%
\bibitem [{\citenamefont {Audi}\ \emph {et~al.}(2003)\citenamefont {Audi},
  \citenamefont {Wapstra},\ and\ \citenamefont {Thibault}}]{Aud03}%
  \BibitemOpen
  \bibfield  {author} {\bibinfo {author} {\bibfnamefont {G.}~\bibnamefont
  {Audi}}, \bibinfo {author} {\bibfnamefont {A.~H.}\ \bibnamefont {Wapstra}}, \
  and\ \bibinfo {author} {\bibfnamefont {C.}~\bibnamefont {Thibault}},\ }\href
  {\doibase 10.1016/j.nuclphysa.2003.11.003} {\bibfield  {journal} {\bibinfo
  {journal} {Nucl. Phys.}\ }\textbf {\bibinfo {volume} {A729}},\ \bibinfo
  {pages} {337} (\bibinfo {year} {2003})}\BibitemShut {NoStop}%
\bibitem [{\citenamefont {Suzuki}\ \emph {et~al.}(2003)\citenamefont {Suzuki},
  \citenamefont {Sagawa},\ and\ \citenamefont {Hagino}}]{Suz03}%
  \BibitemOpen
  \bibfield  {author} {\bibinfo {author} {\bibfnamefont {T.}~\bibnamefont
  {Suzuki}}, \bibinfo {author} {\bibfnamefont {H.}~\bibnamefont {Sagawa}}, \
  and\ \bibinfo {author} {\bibfnamefont {K.}~\bibnamefont {Hagino}},\ }\href
  {\doibase 10.1103/PhysRevC.68.014317} {\bibfield  {journal} {\bibinfo
  {journal} {Phys. Rev. C}\ }\textbf {\bibinfo {volume} {68}},\ \bibinfo
  {pages} {014317} (\bibinfo {year} {2003})}\BibitemShut {NoStop}%
\bibitem [{\citenamefont {Elekes}\ \emph {et~al.}(2005)\citenamefont {Elekes}
  \emph {et~al.}}]{Ele05}%
  \BibitemOpen
  \bibfield  {author} {\bibinfo {author} {\bibfnamefont {Z.}~\bibnamefont
  {Elekes}} \emph {et~al.},\ }\href {\doibase 10.1016/j.physletb.2005.04.007}
  {\bibfield  {journal} {\bibinfo  {journal} {Phys. Lett. B}\ }\textbf
  {\bibinfo {volume} {614}},\ \bibinfo {pages} {174} (\bibinfo {year}
  {2005})}\BibitemShut {NoStop}%
\bibitem [{\citenamefont {{Kobayashi}}\ \emph {et~al.}(2012)\citenamefont
  {{Kobayashi}} \emph {et~al.}}]{Kob12}%
  \BibitemOpen
  \bibfield  {author} {\bibinfo {author} {\bibfnamefont {N.}~\bibnamefont
  {{Kobayashi}}} \emph {et~al.},\ }\href {\doibase 10.1103/PhysRevC.86.054604}
  {\bibfield  {journal} {\bibinfo  {journal} {Phys. Rev. C}\ }\textbf {\bibinfo
  {volume} {86}},\ \bibinfo {pages} {054604} (\bibinfo {year}
  {2012})}\BibitemShut {NoStop}%
\bibitem [{\citenamefont {Urata}\ \emph {et~al.}(2011)\citenamefont {Urata},
  \citenamefont {Hagino},\ and\ \citenamefont {Sagawa}}]{Ura11}%
  \BibitemOpen
  \bibfield  {author} {\bibinfo {author} {\bibfnamefont {Y.}~\bibnamefont
  {Urata}}, \bibinfo {author} {\bibfnamefont {K.}~\bibnamefont {Hagino}}, \
  and\ \bibinfo {author} {\bibfnamefont {H.}~\bibnamefont {Sagawa}},\ }\href
  {\doibase 10.1103/PhysRevC.83.041303} {\bibfield  {journal} {\bibinfo
  {journal} {Phys. Rev. C}\ }\textbf {\bibinfo {volume} {83}},\ \bibinfo
  {pages} {041303} (\bibinfo {year} {2011})}\BibitemShut {NoStop}%
\bibitem [{\citenamefont {Moro}\ and\ \citenamefont {Lay}(2012)}]{Mor12a}%
  \BibitemOpen
  \bibfield  {author} {\bibinfo {author} {\bibfnamefont {A.~M.}\ \bibnamefont
  {Moro}}\ and\ \bibinfo {author} {\bibfnamefont {J.~A.}\ \bibnamefont {Lay}},\
  }\href {\doibase 10.1103/PhysRevLett.109.232502} {\bibfield  {journal}
  {\bibinfo  {journal} {Phys. Rev. Lett.}\ }\textbf {\bibinfo {volume} {109}},\
  \bibinfo {pages} {232502} (\bibinfo {year} {2012})}\BibitemShut {NoStop}%
\end{thebibliography}%

\end{document}